\documentclass[twocolumn,prb,aps,floatfix,longbibliography,superscriptaddress]{revtex4-2}
\pdfoutput=1
\usepackage{color}
\usepackage{graphicx}
\usepackage{physics}
\usepackage{amsthm}
\usepackage{amsmath}
\usepackage{amssymb}
\usepackage{enumerate}
\usepackage{placeins}
\usepackage{booktabs}
\usepackage{dsfont}
\usepackage{hyperref}

\newcommand{\kB}{k_{\mathrm{B}}}

\usepackage{siunitx}
\DeclareSIUnit\angstrom{\text{\AA}}
\AtBeginDocument{%
\heavyrulewidth=.08em
\lightrulewidth=.05em
\cmidrulewidth=.03em
\belowrulesep=.65ex
\belowbottomsep=0pt
\aboverulesep=.4ex
\abovetopsep=0pt
\cmidrulesep=\doublerulesep
\cmidrulekern=.5em
\defaultaddspace=.5em
}

\usepackage[color=orange!60,textsize=scriptsize]{todonotes}
\usepackage{pdfpages} % include pdfs
\usepackage{pgffor} % for loops
\usepackage{xr} % referencing the supplement

\makeatletter
\AtBeginDocument{\let\LS@rot\@undefined}
\makeatother

\def\supplementfilename{supplement}

\newif\ifarXiv
\arXivfalse

\ifarXiv
    \pdfximage{\supplementfilename.pdf}
    \def\numbersupplementpages{\the\pdflastximagepages}
\fi

\begin{document}

\title{Interface Engineering of Helium Confinement in Argon-Preplated MCM-41 Nanopores}

\author{Rahul Soni}
\thanks{These authors contributed equally to this work.}
\affiliation{Department of Physics, Indiana University, Bloomington, IN 47405, USA}
\author{Nathan S. Nichols}
\thanks{These authors contributed equally to this work.}
\affiliation{Argonne Leadership Computing Facility, Argonne National Laboratory, Lemont, IL 60439, USA}
\author{Sutirtha Paul}
\affiliation{Department of Physics and Astronomy, University of Tennessee, Knoxville, TN 37996, USA}
\author{Garfield Warren}
\affiliation{Department of Physics, Indiana University, Bloomington, IN 47405, USA}
\author{Paul Sokol}
\affiliation{Department of Physics, Indiana University, Bloomington, IN 47405, USA}
\author{Adrian Del Maestro}
\affiliation{Department of Physics and Astronomy, University of Tennessee, Knoxville, TN 37996, USA}
\affiliation{Min H. Kao Department of Electrical Engineering and Computer Science, University of Tennessee, Knoxville, TN 37996, USA}

\begin{abstract}
Atomic-scale modification of mesopore interfaces provides a route to tune the confinement experienced by adsorbed fluids, but how a specific interface preparation translates into the resulting microscopic confinement potential remains unclear. Here, we show that preplating MCM-41 with an argon monolayer modifies the effective pore interface by occupying strongly attractive regions of the heterogeneous silica surface and screening its atomic-scale corrugation. Grand-canonical Monte Carlo simulations of argon adsorption, low-temperature molecular dynamics, and helium test-particle insertion are combined with adsorption isotherms and neutron-scattering measurements to characterize the preplated pore at the atomic scale. Helium test-particle insertion calculations show that the modified interface shifts the helium adsorption minimum to an annular region inside the pore and produces a confinement landscape dominated by a smooth radial component.  The resulting radial confinement potential can be described by a continuum cylindrical model, providing microscopic support for the effective potential used in earlier quantum Monte Carlo studies. Residual corrugation persists over multiple spatial scales and is accurately captured by a Gaussian process surrogate. These results demonstrate how atomic preplating can tailor nanopore confinement and provide an experimentally constrained microscopic potential for predictive studies of confined quantum fluids.
\end{abstract}

\maketitle
% ------------------------------------------------------------------------------
% ==============================================================================

\section{Introduction}

Ordered mesoporous silica is not merely a geometric container: local surface chemistry and atomic-scale topology conspire to create heterogeneous adsorption landscapes that control the structure and dynamics of confined fluids. Experiment-guided models of MCM-41 \cite{Kresge:1992kd} show that pore symmetry, curvature, and chemically inequivalent surface sites materially alter adsorption and selectivity \cite{Ugliengo:2008ks,Andrea:2022,Carta:2023} while pore-wall chemistry can reorganize a nanoconfined fluid \cite{Weinberger:2022}. At low temperatures, the confinement of light atoms such as helium with large zero-point motion into MCM-41 nanopores has enabled the study of dimensional crossover on superfluidity \cite{Wada:2001jb,Wada:2005uc,Ikegami:2005ec,Toda:2007cv,Taniguchi:2011bx,Taniguchi:2013us,Yager:2013cv,Demura:2015hq,Demura:2017gy,Taniguchi:2018ip,Prisk:2013cu,Bryan:2017hb,Bryan:2018vb,Bossy:2019qd,Taniguchi:2020ln,Huan:2020ya}. Pushing these systems toward narrower effective pore geometries is especially appealing, since sufficiently narrow confinement may provide access to more distinctly quasi-one-dimensional regimes, including the possible realization of Tomonaga-Luttinger-liquid behavior in helium, a type of emergent quantum hydrodynamics \cite{DelMaestro:2011,DelMaestro:2022}. 

A major practical obstacle, however, is that mesoporous materials as synthesized often have pores that are too wide, too structurally complex, or too strongly adsorbing to produce ideal and tunable quasi-one-dimensional energetic confinement. A promising route to overcome this limitation is to modify the pore environment through rare-gas or alkali metal preplating, for example with argon or cesium \cite{McNamara:2025, Paul:2026}.  Preplating provides a distinct route to interface engineering, as an adsorbed atomic layer can reduce the available pore radius while simultaneously occupying high-affinity surface crevices and defining a new confining surface for a second fluid. 

Earlier quantum Monte Carlo calculations demonstrated that Ar preplating can stabilize an emergent quasi-one-dimensional helium core with tunable density in MCM-41 \cite{Nichols:2020}. To make those large-scale many-body calculations tractable, the heterogeneous preplated pore was deliberately coarse-grained into a smooth, radially symmetric confinement potential informed by measured pore and monolayer properties. This separation of scales isolated the dominant effects of preplating embodied by the assumption that the Ar monolayer screens enough atomic-scale silica roughness for the confined helium to experience approximately cylindrical confinement. 

In this paper, we test and resolve this hypothesis by constructing the helium confinement potential from an atomistically resolved model of an experimentally characterized Ar-preplated MCM-41 pore.  Building on an experiment-guided atomistic model of hexagonal MCM-41 \cite{Andrea:2022}, grand-canonical Monte Carlo and low-temperature molecular dynamics are combined to reconstruct the Ar monolayer and benchmark its thermodynamics and structure against adsorption and neutron-scattering measurements.  Helium test-particle insertion then yields the spatially resolved one-body potential, revealing that the Ar monolayer does more than narrow the pore: it fills the most strongly corrugated regions of the silica surface and transforms its highly heterogeneous adsorption landscape into an annular minimum at $r \simeq \SI{11.7}{\angstrom}$ with a depth of approximately $\SI{44}{\kelvin}$. Remarkably, the radial average of this atomistic potential is accurately described by the smooth cylindrical form used in Ref.~\cite{Nichols:2020}, providing direct microscopic validation of the approximation that enabled those quantum Monte Carlo calculations. At the same time, we introduce a Gaussian process surrogate model to resolve the residual corrugation across multiple length scales, achieving a sub-kelvin mean reconstruction error. The result is an experimentally constrained, atomistically resolved confinement potential that links the preparation of a real nanopore with a preplated Ar interface directly to the one-body term in a quantum many-body Hamiltonian.

The remainder of this paper is organized as follows. We first summarize the experimental characterization of the MCM-41 sample and the neutron-scattering measurements used to benchmark the simulations. We then describe the atomistic MCM-41 model and the simulation methodology used for argon adsorption. Next, we present results for adsorption, radial structure, and static correlations of the argon monolayer, together with comparisons to experiment. Finally, we analyze helium test-particle insertion in the preplated pore and discuss the implications for the smoothness and residual corrugation of the confinement potential.

% ------------------------------------------------------------------------------
% ==============================================================================

\section{Experimental Details}
\label{sec:experimental_details}

The MCM-41 sample was obtained from Sigma-Aldrich \cite{mcm41SA:2008} and was characterized using X-ray powder diffraction and N$_2$ gas adsorption isotherm measurements \cite{Prisk:2013cu}. The X-ray diffraction data indicated that the sample consisted of a single phase with pores arranged on a hexagonal lattice with a lattice constant of \SI{4.7}{\nano\meter}.  A Brunauer-Emmett-Teller (BET) analysis \cite{Brunauer:1938pz,Schlumberger:2021,Thommes:2015IUPAC} of the N$_2$ isotherm gave a surface area of \SI{915}{\meter^2\per\gram}.  The pore diameter size distribution was calculated using the Kruk-Jaroniec-Sayari method \cite{Jaroniec:1999mi} and was found to be Gaussian with a mean value of \SI{3.0}{\nano\meter} and a full-width at half-maximum of \SI{0.3}{\nano\meter}. 

Adsorption isotherms measured with research-grade Ar at $T = \SI{90}{\kelvin}$ were used to determine monolayer coverage where a BET analysis gave \SI{8.994}{\milli\mol\per\gram} \cite{Nichols:2020}. Together with the measured specific surface area above, this corresponds to an areal number density of $\SI{0.059}{\angstrom^{-2}}$.  Approximating the effective monolayer thickness by the Lennard–Jones size $\sigma_{\text{Ar}} = \SI{3.405}{\angstrom}$ yields a monolayer density of $n_{\rm Ar} = \SI{0.017}{\angstrom^{-3}}$.

Neutron scattering studies of $^4$He in Ar preplated MCM-41 were carried out using the Disk Chopper Spectrometer (DCS) at the NIST Center for Neutron Research \cite{Copley:2003dc}. This instrument is a direct geometry time-of-flight chopper spectrometer which views a cold moderator. High speed choppers are used to create a pulsed neutron beam with a well defined incident wavelength. Neutrons scattered by the sample are detected by a secondary spectrometer consisting of 913 $^3$He detectors \SI{4.01}{\meter} from the sample at scattering angles from \SI{5}{\degree} to \SI{140}{\degree}. Standard data reduction routines \cite{Azuah:2009cs} were used to convert the observed scattering to the dynamic structure factor $S(q,E)$.

The sample for these studies consisted of \SI{6.13}{\gram} of MCM-41 inside a cylindrical aluminum can.  The MCM-41 was in the form of cylindrical pellets \SI{1.25}{\centi\meter} in diameter and \SI{1}{\centi\meter} high with a mass of \SI{0.875}{\gram}.  The pellets were baked in vacuum at \SI{120}{\celsius} to remove adsorbed water vapor.  The sample was then transferred to an aluminum sample cell in a nitrogen glove box.  The cell was a cylindrical aluminum can of outer diameter \SI{1.5}{\centi\meter}, a height of \SI{6}{\centi\meter}, and a wall thickness of \SI{1}{\milli\meter}. The pellets were separated by cadmium spacers to reduce multiple scattering. A top-loading liquid helium cryostat with aluminum tails, commonly referred to as an ``orange" cryostat, was used to obtain the low temperatures examined in this study with a silicon diode used to monitor the sample temperature.

Incident wavevectors of \SI{4.0}, \SI{2.5} and \SI{1.71} \AA$^{-1}$ were used allowing momentum transfers, $q$, from \SI{0.70} to \SI{6.95} \AA$^{-1}$ to be studied.  The dynamic structure factor, $S(q,E)$, was measured for both the empty MCM-41 and with a monolayer coating.  The static structure factor, $S(q)$ was obtained by integrating $S(q,E)$ over the energy.  The empty MCM-41 scattering was then subtracted from the MCM-41 with a monolayer of Ar.  The resultant scattering captures the Ar-Ar and Ar-MCM-41 correlations and is compared with predictions from numerical simulations in the Results section. 

\begin{figure}[t]
    \centering
    \includegraphics[width=\columnwidth]{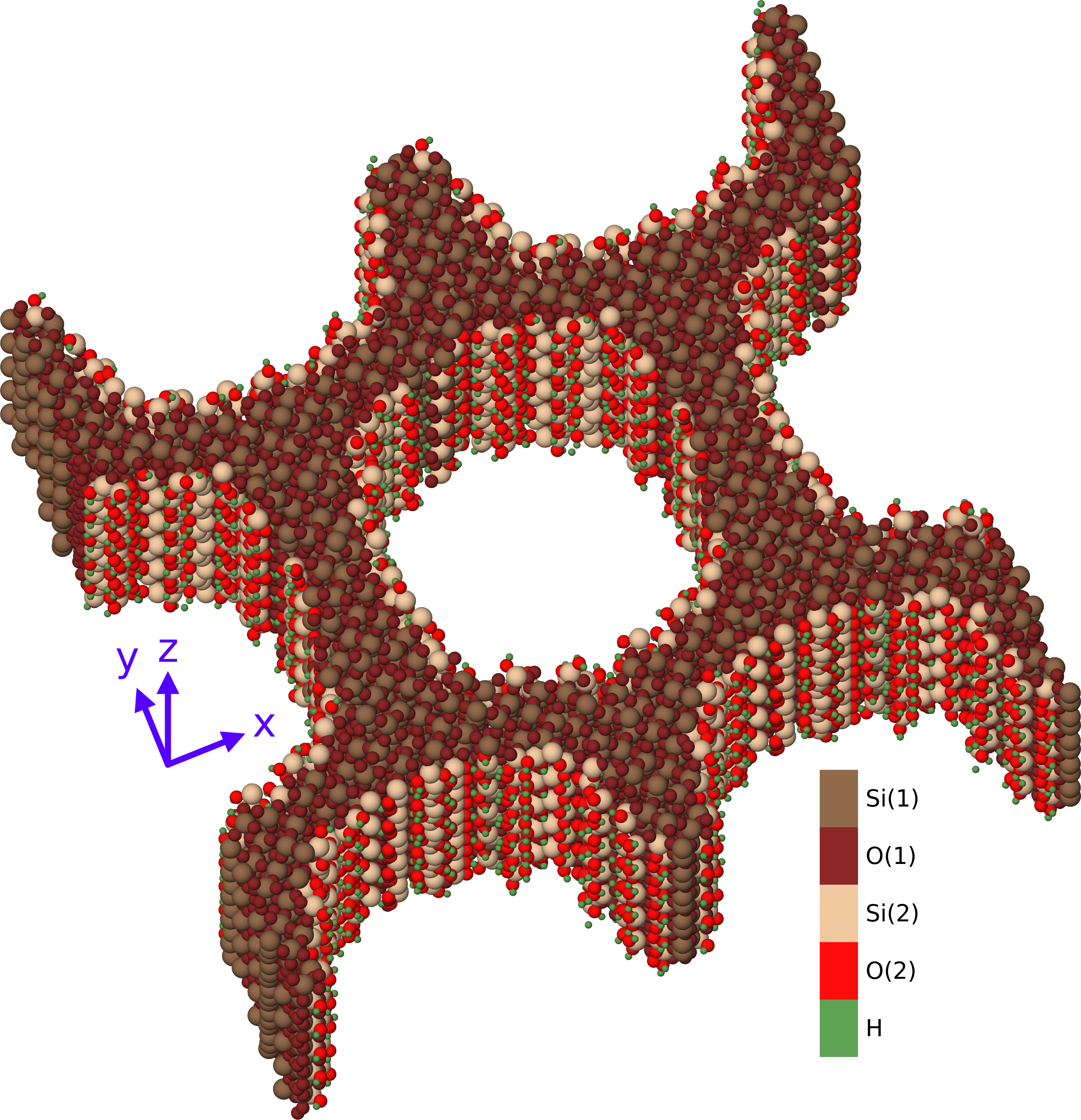}
    \caption{\label{fig:MCM-41 structure} Perspective view of the atomistic MCM-41 supercell used in this work composed of $2\times 1\times 5$ primitive cells forming a nearly cylindrical pore extended along the $z$-direction. The legend indicates the different atom types in the MCM-41 structure (see text for discussion).}
\end{figure}

\section{Simulation Details}\label{sec:methods}
\subsection{Grand-canonical Monte Carlo Adsorption}\label{ssec:gcmc_methods}

Grand-canonical Monte Carlo (GCMC) simulations of argon adsorption inside a single MCM-41 pore at $T=\SI{90}{\kelvin}$ were performed using LAMMPS~\cite{thompson:2022,plimpton:1995}.  The silica framework was taken from the experiment-guided atomistic hexagonal MCM-41 model developed in Ref.~\cite{Andrea:2022} in which Si(1)/O(1) corresponds to bulk $\mathrm{SiO}_2$ atoms and Si(2)/O(2) denote pore surface atoms.  Subsequent calculations with this model show that curvature and chemically inequivalent surface sites materially affect adsorption \cite{Carta:2023}, motivating retention of the atomistic pore surface here.  Interactions were modeled using Lennard-Jones (LJ) potentials with parameters listed in Table~\ref{tab:LJ_parameters}, where vanishing LJ parameters were assigned to the bulk-like Si atoms since their direct role in gas adsorption is  negligible~\cite{yun:2002,furukawa:2005}. Cross interactions between different atomic species were generated using Lorentz-Berthelot (LB) mixing rules \cite{1964:HirschfelderBook} in LAMMPS. A single pore simulation cell was constructed by replicating the unit cell twice along the $x$-direction and five times along the $z$-direction, giving a length of $L_z = \SI{107.49}{\angstrom}$. The resulting MCM-41 cylindrical pore structure considered in this work is shown in Fig.~\ref{fig:MCM-41 structure}, with the pore centered at $(x_0,y_0)=(\SI{42.996}{\angstrom},\SI{39.413}{\angstrom})$. The minimum distance from the pore center to the silica framework is $R_{\rm min}=\SI{16.064}{\angstrom}$, consistent with the measured radius in Sec.~\ref{sec:experimental_details}. For the purpose of argon adsorption analysis, in this study we adopt a larger nominal pore radius of $R_{\rm pore}=\SI{20}{\angstrom}$ so as to encompass the full pore region, including the local crevices and surface corrugation of the atomistic silica wall.

% We specifically employ the DFT-based \(P6mm\) representation of MCM-41, since the canonical hexagonal \(P6mm\) symmetry is the experimentally relevant one for MCM-41, and among the models examined in Ref.~\cite{andrea:2022} the MCM-41/DFT structure was identified as the most realistic representation, providing the correct space-group symmetry, the DFT-derived pore size, the closest match to the experimentally constrained surface chemistry, and the best agreement with the adsorption isotherm. 

\begin{table}
    \renewcommand{\arraystretch}{1.5}
    \setlength\tabcolsep{12pt}
    \begin{tabular}{@{}lll@{}} 
        \toprule
        Atom & $\sigma\; \qty[\si{\angstrom}]$ & $\varepsilon/\kB\; \qty[\si{\kelvin}]$ \\
        \midrule
        Ar      & 3.405 & 119.8 \\
        Si(1)   & 0.0  & 0.0 \\
        O(1)    & 2.708  & 184.979 \\
        Si(2)   & 3.905  & 59.415 \\
        O(2)    & 3.07  & 120.032\\
        H       & 0.0  & 0.0 \\
        \bottomrule
    \end{tabular}
    \caption{\label{tab:LJ_parameters} Lennard-Jones parameters taken from Ref.~\cite{Andrea:2022} used in the simulations of argon in MCM-41 nanopores. (1) refers to bulk and (2) to pore surface atoms.}
\end{table}

% The GCMC simulations utilized periodic boundary conditions along the pore axis $z$ and open boundary conditions along the $x$- and $y$-directions. Chemical potentials were scanned over the range $\mu=-17.00,{\rm kJ/mol}$ to $-9.00,{\rm kJ/mol}$ in steps of $\Delta\mu=0.05,{\rm kJ/mol}$, while the MCM-41 framework was treated as rigid throughout the simulations. For quantities directly compared with experiment, including the adsorption isotherm and static structure factor, five independent GCMC simulations with different random seeds were performed to improve statistical reliability. For the remaining calculations, including representative radial density profiles, pair-correlation functions, and configurations used for subsequent MD relaxation, a single random seed was used. For each value of the chemical potential, the system was first equilibrated for $200000$ Monte Carlo steps, followed by a production run of $800000$ steps. Atomic configurations and particle-number statistics generated during the production stage were used to compute adsorption properties, radial density profiles, and static correlation functions discussed below. For the isotherm and structure-factor comparisons, observables were first averaged over production snapshots for each random seed and then averaged over the five independent seeds; 

The GCMC simulations utilized periodic boundary conditions along the pore axis $z$ and open boundary conditions along the $x$- and $y$-directions. The chemical potential $\mu$ was scanned over the range 
$\mu = \SIrange{-17.0}{-9.0}{\kilo\joule\per\mol}$
in steps of $\Delta\mu=\SI{0.05}{\kilo\joule\per\mol}$, while the MCM-41 framework was treated as rigid throughout the simulations. To improve statistical reliability, five independent GCMC simulations with different random seeds were performed. For each value of the chemical potential, the system was first equilibrated for $2\times 10^5$ Monte Carlo steps, followed by a production run of $8\times 10^5$ steps. Atomic configurations and particle number statistics generated during the production stage were used to compute adsorption properties, radial density profiles, and static correlation functions discussed below. For each random seed, these quantities were first averaged over multiple snapshots sampled during the production run, and the final reported results were then obtained by averaging over the five independent seeds employing standard error analysis techniques.

\subsection{Adsorption and Structural Observables}
\label{ssec:observables}

Adsorption properties can be studied by monitoring the average number of argon atoms inside the pore as a function of chemical potential. We calculate the adsorption uptake which provides the most direct measure of pore filling and allows for a direct comparison with adsorption isotherms measured in experiment. The uptake is expressed as
\begin{equation}
{\rm Uptake}=\frac{\langle N\rangle}{m_{\rm pore}N_A},
\label{eq:uptake}
\end{equation}
where $\langle N\rangle$ is the average number of adsorbed argon atoms, $m_{\rm pore}$ is the mass of the MCM associated with a single pore, calculated by dividing the total mass of the MCM structure by the number of pores, and $N_A$ is Avogadro's number. Here, $\langle\cdots\rangle$ denotes an average over all independent GCMC random seeds.

To analyze the structural properties of the adsorbed argon layer inside the pore, we computed the average radial density profile measured with respect to the central pore axis. For a configuration containing $N$ argon atoms at positions $\mathbf{r}_i$, the cylindrically symmetric radial density is defined as
\begin{equation}
\rho_{\rm rad}(r)= \frac{1}{2\pi r L_z}\left\langle \sum_{i=1}^{N}\delta\left(r_i-r\right)\right\rangle,
\label{eq:rhorad}
\end{equation}
where $r_i=\sqrt{(x_i-x_0)^2+(y_i-y_0)^2}$ measures the distance from the pore center, $L_z$ is the pore length along the $z$-direction and $N$ is the instantaneous (configuration dependent) number of argon atoms in the GCMC simulations. The evolution of $\rho_{\rm rad}(r)$ as the chemical potential is varied provides a direct picture of the formation of successive adsorption layers inside the pore. To identify monolayer completion, we use these radial density profiles to define the boundary between the first and second adsorbed layers through the local minimum between the corresponding peaks, and then monitor the first-layer occupancy $N_1(\mu)$ and its derivative $dN_1/d\mu$ as functions of chemical potential. Further details are provided in Appendix~\ref{ssec:monolayer_identification}.

To fully characterize spatial correlations within the adsorbed argon monolayer, we compute a cylinder-corrected pair-correlation function for Ar. For each sampled configuration, only argon atoms lying inside the cylindrical pore region of radius $R_{\rm pore}$ are retained, and all distinct pairs $(i<j)$ are used to construct a histogram of pair separations
\begin{equation}
r_{ij}=\sqrt{(x_i-x_j)^2+(y_i-y_j)^2+\tilde z_{ij}^{\,2}},\label{eq:min_img_dist}
\end{equation}
where $\tilde z_{ij}$ denotes the minimum-image separation along the periodic pore axis $z$. The resulting pair histogram $H(r)$ is then normalized by a reference histogram $H_{\rm ideal}(r)$ obtained from uniformly distributed particles inside the same cylindrical geometry, with the same number of particles and $L_z$ for each configuration. The pair-correlation function is therefore defined as
\begin{equation}\label{eq:pair_corr_function}
g_{\rm cyl}(r)=\frac{H(r)}{H_{\rm ideal}(r)}
\end{equation}
where the normalization removes geometric effects associated with cylindrical confinement and finite system size. This provides a more meaningful measure of local structure in the preplated argon layer than a conventional bulk spherically normalized radial distribution function.

For direct comparison with neutron-scattering measurements, we also compute the powder-averaged static structure factor from the atomistic GCMC configuration using the Debye formula~\cite{Debye:1915}. Experimentally, the scattering from empty MCM-41 is subtracted from that of the Ar-preplated sample. The MCM-MCM contribution is therefore removed, leaving the Ar-Ar contribution and the cross correlations between the adsorbed Ar atoms and the MCM-41 framework. Thus, we model the corresponding loaded-minus-empty structure factor as
\begin{align}
S(q) &= S_{\rm Ar-Ar}(q) + \sum_{\beta\in\mathcal{M}} S_{\rm Ar-\beta}(q) \notag\\
&= \frac{\displaystyle\sum_{i,j\in{\rm Ar}} b_i b_j \frac{\sin(qr_{ij})}{qr_{ij}}}{ \displaystyle \sum_{i\in{\rm Ar}}b_i^2} + \sum_{\beta\in\mathcal{M}} \frac{ \displaystyle 2\sum_{\substack{i\in{\rm Ar}\\ j\in\beta}} b_i b_j\frac{\sin(qr_{ij})}{qr_{ij}} }{ \displaystyle \sqrt{ \left(\sum_{i\in{\rm Ar}}b_i^2\right) \left(\sum_{j\in\beta}b_j^2\right) }},
\label{eq:SQ}
\end{align}
where $\mathcal{M}=\{\mathrm{Si(1)},\mathrm{O(1)},\mathrm{Si(2)},\mathrm{O(2)},\mathrm{H}\}$ is the set of MCM-41 framework species atoms, $q$ is the magnitude of the scattering wave vector, $r_{ij}$ is the minimum image distance between atoms $i$ and $j$ as in Eq.~\eqref{eq:min_img_dist}, and $b_i$ is the coherent neutron scattering length of atomic species $i$. 

The first term describes correlations within the adsorbed Ar layer, whereas the second term contains the cross correlations between the Ar atoms and the MCM-41 framework atoms. The factor of two in the cross term accounts for the equivalent Ar-MCM and MCM-Ar contributions to the full double sum. The resulting $S(q)$ therefore isolates the scattering associated with the adsorbed Ar layer and its correlations with the pore surface, providing a direct quantity for comparison with experiment. 
%For the calculation of cross correlated structure factor, framework atoms such as bulk Si(1) and H are ignored (see Table~\ref{tab:LJ_parameters}).

\subsection{Molecular Dynamics and Test Particle Insertion}\label{ssec:md_tpi}

While the GCMC simulations provide the adsorption and structural properties of the argon layer at $T=\SI{90}{\kelvin}$, the resulting configurations are not representative of the quenched low-temperature preplated pores studied experimentally.  Thus, we use GCMC configurations near monolayer completion and perform a molecular dynamics (MD) relaxation in LAMMPS while keeping the MCM-41 framework rigid. The subsequent MD evolution at $T=\SI{90}{\kelvin}$, followed by cooling to $T=\SI{4}{\kelvin}$, allows the argon monolayer to energetically reorganize within the pore and yields an effectively frozen Ar-preplated configuration that generates a static one-body confinement potential experienced by helium atoms inside the pore.  

We use these frozen Ar-preplated MCM-41 configurations to perform test particle insertion (TPI) calculations,  where a single He atom is placed at multiple points spanning the pore volume and the interaction energy between the probe atom and the frozen configuration is evaluated at each point. For a given frozen host configuration $\alpha$, the resulting interaction energy at probe position $\mathbf{r}$ is given by
\begin{equation}
    U_{\alpha}(\mathbf{r}) = \sum_{j\in {\rm host}}u_{\vphantom{i^i}{\rm He},s_j}\left(|\mathbf{r} - \mathbf{R}_j^{\alpha}|\right)
\end{equation}
where $\mathbf{R}_{j}^{\alpha}$ denotes the position of host atom $j$ in frozen host configuration $\alpha$, $s_j$ denotes its atomic species, and $u_{{\rm He},s_j}$ is the pair interaction between the He probe and the host species $s_j$. Lennard-Jones interaction parameters used for He TPI simulations are $\sigma=\SI{2.640}{\angstrom}$, $\varepsilon/k_B=\SI{10.9}{\kelvin}$.  Note that the host now comprises both the preplated Ar atoms and the MCM-41 framework atoms. In cylindrical coordinates, $U_{\alpha}(\mathbf{r})\equiv U_{\alpha}(r,\theta,z)$, provides a spatially resolved one-body confinement potential experienced by helium inside the preplated pore. From this quantity, we can probe microscopic structure and construct both transverse potential maps and effective radial confinement potentials relevant for subsequent low-temperature helium modeling.

\subsection{Interpolation with Gaussian process regression}\label{ssec:gp}
To construct a continuous potential energy surface for the z-averaged planar potential from the discrete set of points obtained via TPI we use Gaussian Process (GP) regression. This provides a flexible, nonparametric framework for constructing smooth approximations to functions based on discrete training data \cite{Rasmussen:2005gp} and has been shown to be a powerful tool for modeling non-covalent potential energy surfaces \cite{Kolb2017,Uteva2018,Wiens2019MFGP,Deringer:2019AdvMater,Deringer2021ky,Schneider2023,Akram2026}.

Mathematically, given a set of points for which the potential energy, V is known $\mathcal{D} = \qty{\vb*{R}, \vb*{U}} = \qty{(\vb*{r}_i, U_i)}_{i=1}^n,\quad \text{where } U_i = U(\vb*{r}_i)$, we can model the collection of energy values as a multivariate normal distribution,
 $\vb*{U} \sim \mathcal{N}\qty(\bar{U}, \mathsf{K}(\vb*{R},\vb*{R}))$
where the covariance matrix $\mathsf{K}$ has elements
$ \mathsf{K}_{ij} = k(\vb*{r}_i,\vb*{r}_j) + \sigma_U^2 \delta_{ij}$. A small noise variance, $\sigma_U^2$,  accounts for numerical uncertainty or residual mismatch between the data and the model, $\delta_{ij}$ is the Kronecker delta and $\bar{U}_i = \mathbb{E}\bqty{U(\vb*{r}_i)}$ are the components of a (possible) mean function across the dataset. The kernel function $k(\vb*{r}_i,\vb*{r}_j)$, captures correlations of spatially proximate interaction energies between two positions $\vb*{r}_i$ and $\vb*{r}_j$ and can be chosen to best capture the properties of a given target function. 

Within the GP framework, the potential energy at a new position $\vb*{r}_{n+1}$ not contained in the dataset is given by:
\begin{equation}
    U_{n+1} = \bar{U}(\vb*{r}_{n+1}) + \vb*{k}(\vb*{r}_{n+1})^\top \mathsf{K}^{-1} \bqty{\vb*{U} - \bar{U}(\vb*{R})} \label{Eq:posteriormean}\\
\end{equation}
where the kernel vector $\vb{k}$ is given by $
\vb{k}(\vb*{r}_{n+1}) =
\qty[k(\vb*{r}_1,\vb*{r}_{n+1}), \ldots, k(\vb*{r}_n,\vb*{r}_{n+1})]^{\top}.$
The mean function and the hyperparameters of the kernel function are obtained from the training data by maximizing the log-marginal-likelihood,
$
\mathcal{L} = \ln \qty[(2\pi)^{-n/2} \abs{\mathsf{K}}^{-1/2} \exp(-\frac{1}{2}(\vb*{U} -  \bar{U})^\top \mathsf{K}^{-1}(\vb*{U} -  \bar{U}))]$. 

\section{Results}\label{sec:results}
\subsection{Thermodynamics and Structure}\label{ssec:GCMC_results}

\begin{figure}[t]
    \centering
    \includegraphics[width=\columnwidth]{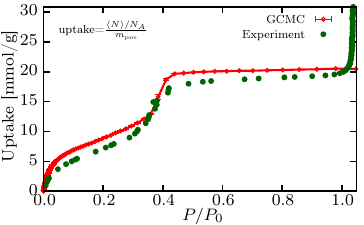}
    \caption{\label{fig:uptake_vs_pressure} Adsorption isotherm of argon inside a MCM-41 nanopore measured via experiment at $T=\SI{90}{\kelvin}$ compared with grand canonical Monte Carlo. Pressures have been scaled to the monolayer filling inflection point and we find $P_0 = \SI{2.6}{\bar}$.}
\end{figure}

We begin by examining the adsorption isotherm obtained from GCMC simulations measured via Eq.~\eqref{eq:uptake}. Figure~\ref{fig:uptake_vs_pressure} shows the argon uptake in the MCM-41 pore as a function of normalized pressure $P/P_0$, together with the corresponding experimental data. 
To place the simulation results on the same horizontal scale as experiment, the chemical potential used in the GCMC calculations was converted to a reduced pressure through
\begin{equation}
\frac{P}{P_0}=\frac{P_{\rm ref}}{P_0}\exp\!\left[\frac{\mu-\mu_{\rm ref}}{R_g T}\right],
\end{equation}
where $R_g$ is the gas constant, $\mu_{\rm ref}=\SI{-9.85}{\kilo\joule\per\mol}$ is a reference chemical potential determined from the inflection point of the GCMC adsorption isotherm and $P_{\rm ref}/P_0 = 0.38$ was identified from the same quantity in the experimental data.    At low pressure, the number of Ar atoms in the pore increases gradually as argon begins to adsorb on the pore surface. With increasing $P/P_0$, a more rapid rise in uptake is observed, signaling the buildup of the adsorbed argon layer on the pore wall. Over a broad range, the simulated isotherm is in excellent agreement with the experimental data, indicating that the atomistic pore model and LJ description capture the essential energetics governing argon adsorption in MCM-41 and the high quality of the MCM-41 model in Ref.~\cite{Andrea:2022}. 
At higher pressures, both curves approach a condensation plateau associated with filling of the entire pore \cite{CychoszStruckhoff:2020}. Beyond the condensation plateau, experimentally the space between the grains of the powdered sample gets filled as depicted by the sharp vertical increase in the isotherm, an effect not possible in our single pore system.

\begin{figure}[t]
    \centering
    \includegraphics[width=\columnwidth]{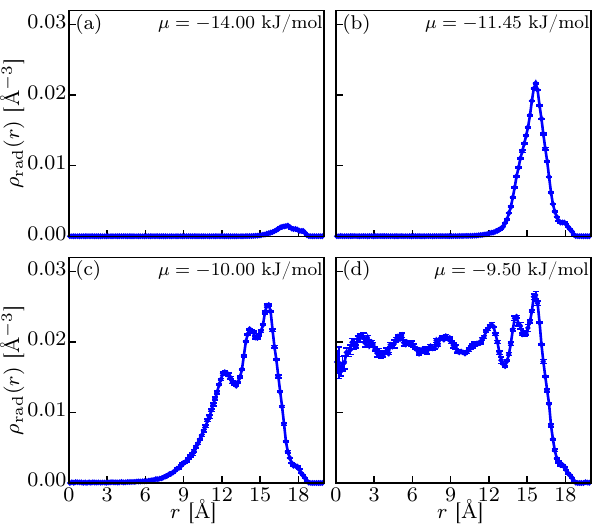}
    \caption{\label{fig:radial_density} The radial density profile of argon inside the MCM-41 pore at four different chemical potentials corresponding to scaled pressures $P/P_0 = \bqty{0.0015,0.0453,0.3148,0.6140}$.
    Panel (a) is in the low-coverage regime, where argon preferentially adsorbs in the crevasses of the MCM wall. Panel (b) shows the formation of a well-defined monolayer regime. Panel (c) corresponds to the multilayer adsorption regime prior to the sharp adsorption jump. Panel (d) is taken just beyond the jump and shows that the pore has become essentially filled with bulk-like argon.}
\end{figure}

With atomistic data over a range of fillings, we next examine the radial density profile, defined in Eq.~\eqref{eq:rhorad} of argon for representative values of the chemical potential, as shown in Fig.~\ref{fig:radial_density}. 
At low chemical potential $\mu=\SI{-14.00}{\kilo\joule\per\mol}$ [panel (a)], the adsorbed argon is confined primarily to the outermost region of the pore, filling crevices and rough portions of the MCM wall with a strongly attractive potential. As the chemical potential is increased to $\mu=-11.45$ kJ/mol [panel (b)], the radial density develops a single sharp peak centered near the pore boundary, signaling the formation of a well-defined monolayer. Upon further increasing the chemical potential to $\mu=-10.00$ kJ/mol [panel (c)], additional peaks emerge at smaller radii, showing that adsorption has progressed beyond the first layer and that multilayer growth has begun. Finally, just beyond the adsorption jump at $\mu=-9.50$ kJ/mol [panel (d)], the density becomes finite throughout the pore interior, consistent with capillary condensation and bulk-like filling of the pore.

\begin{figure}[t]
    \centering
    \includegraphics[width=\columnwidth]{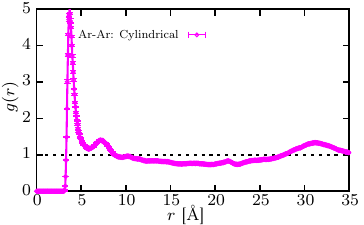}
    \caption{\label{fig:pair_correlation_function} Cylinder corrected Ar-Ar pair correlation function $g(r)$ for the adsorbed monolayer. The sharp first peak depicts the nearest neighbor correlations within the monolayer, the weaker second peak demonstrates the second nearest neighbor correlations, whereas the broader peak at $r\simeq \SI{31.3}{\angstrom}$ reflects the diametric correlations across the cylindrical pore.}
\end{figure}

Figure~\ref{fig:pair_correlation_function} shows the cylinder-corrected Ar-Ar pair-correlation function $g(r)$ as a function of argon particle separation for the monolayer configuration at $\mu=\SI{-11.45}{\kilo\joule\per\mol}$, as calculated by Eq.~\eqref{eq:pair_corr_function}. The first large peak at $r\approx \SI{3.7}{\angstrom}$ corresponds to the nearest neighbor separation within the adsorbed monolayer and is consistent with twice the van der Waals radius of an argon atom. A much weaker secondary peak appears at $r\approx \SI{7.3}{\angstrom}$ that corresponds to second nearest neighbor separation of the Ar atoms within the monolayer. At intermediate distance, the $g(r)$ remains close to unity. However, an additional weaker peak emerges near $r\approx \SI{31.3}{\angstrom}$, capturing the diametric correlations across the cylindrical pore due to the confined geometry of the monolayer.

\begin{figure}[t]
    \centering
    \includegraphics[width=\columnwidth]{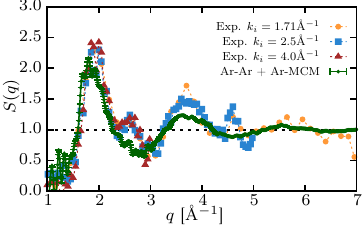}
    \caption{\label{fig:static_structure_factor} Static structure factor obtained from powder averaged Debye formula (in green). The experimental data is obtained from neutron scattering measurements for different incident wavevectors $k_i= \SIlist{1.71;2.5;4.0}{\angstrom^{-1}}$}. 
\end{figure}

A more direct comparison with neutron-scattering measurements is provided by the powder-averaged static structure factor shown in Fig.~\ref{fig:static_structure_factor}. The simulation result was obtained from the Debye formula in Eq.~\eqref{eq:SQ} by combining the Ar-Ar contribution with the Ar-MCM cross correlations, where the latter include the relevant MCM framework atoms near the pore surface up to a cutoff radius of $R_c=\SI{22.5}{\angstrom}$.

The calculated structure factor reproduces many of the main features of the experimental data over range of wave vectors. In particular, the dominant first peak near $q\approx \SI{1.8}{\angstrom^{-1}}$, which reflects the main structural correlations of the adsorbed monolayer, is captured in both position and amplitude. The simulations also capture the broader secondary peak at intermediate $q$, indicating the importance of Ar-MCM cross correlations. Other fine features in $S(q)$ data, including a peak just above $\SI{2.5}{\angstrom^{-1}}$ and persistent modulations at higher $q$, are absent from our simulations. Uncovering the origin of these discrepancies remains an interesting open question. 

\subsection{Adsorption Potential}\label{ssec:TPI_results}

\begin{figure*}[t]
    \centering
    \includegraphics[width=0.9\linewidth]{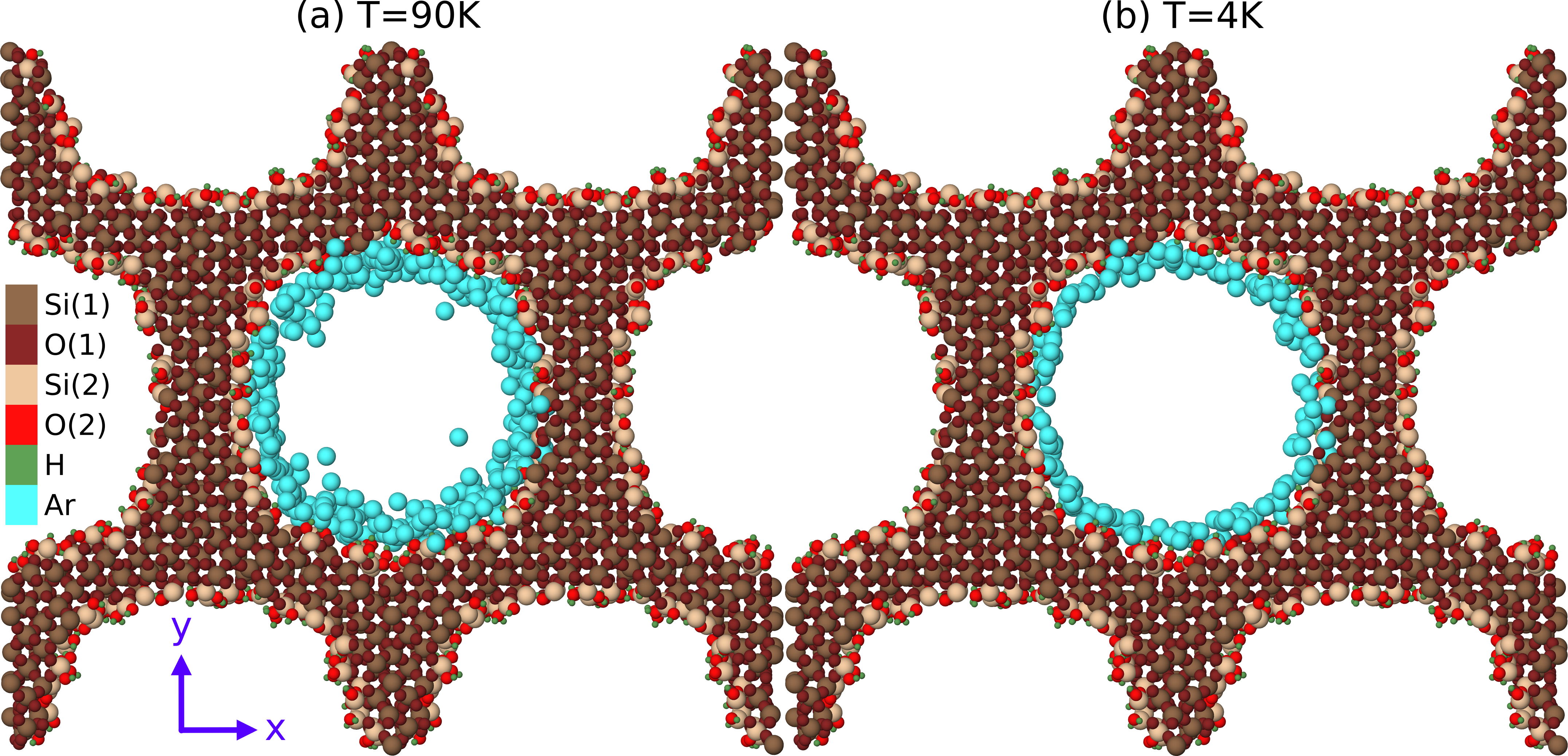}
    \caption{\label{fig:Ar_gcmc_md_snapshots} Top view (along $z$) of Ar inside MCM-41 at monolayer completion. The left panel shows the configuration obtained directly from GCMC simulation at $T=\SI{90}{\kelvin}$, while the right panel shows the configuration after MD relaxation and cooling to $T=\SI{4}{\kelvin}$. Both configurations contain $N=515$ argon atoms. The low temperature quenched configurations are subsequently used as frozen host structures for helium test particle insertion calculations.}
\end{figure*}

Having established the adsorption and structure of Ar in MCM-41, we now turn to the confinement potential experienced by helium in the preplated pore at low temperature. Starting from GCMC configurations at monolayer completion, $\mu=\SI{-11.45}{\kilo\joule\per\mol}$ and $T=\SI{90}{\kelvin}$, we perform molecular-dynamics relaxation followed by cooling to $T=\SI{4}{\kelvin}$ while keeping the MCM-41 framework rigid. Representative configurations before and after cooling are shown in Fig.~\ref{fig:Ar_gcmc_md_snapshots}. The Ar layer already forms a well-defined annular coating of the pore wall at $T=\SI{90}{\kelvin}$, while cooling allows the monolayer to reorganize into a more compact low-temperature configuration without changing the number of Ar atoms in the pore.

\begin{figure}[h]
    \centering
    \includegraphics[width=\columnwidth]{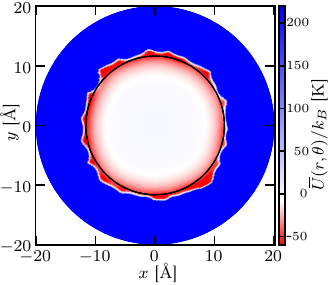}
    \caption{\label{fig:tpi_heatmap} Helium confinement potential (divided by $k_B$) in the Ar-preplated MCM-41 pore at $T=\SI{4}{\kelvin}$, averaged over the pore axis ($z$-direction). The potential minimum forms an annular region, with residual angular corrugation inherited from the atomistic structure of the Ar-preplated interface. The black circle corresponds to the minimum of the radial confinement potential.}
\end{figure}

Using the low-temperature Ar-preplated MCM-41 configurations as frozen host structures, we perform helium test-particle insertion calculations to obtain the corresponding spatially resolved confinement potential. Averaging over ten frozen host realizations obtained from five independent seeds gives the averaged potential $\langle U(r,\theta,z)\rangle$. We further average along the pore axis to define the transverse confinement potential,
\begin{equation}
    \,\overline{\!U}(r,\theta)=\frac{1}{N_z}\sum_{z}\langle U(r,\theta,z)\rangle,
\end{equation}
where $N_z$ is the number of sampled slices along the pore axis. The resulting transverse potential map at $T=\SI{4}{\kelvin}$ is shown in Fig.~\ref{fig:tpi_heatmap}. The heatmap reveals that the most attractive  region forms an approximately annular minimum. The black circle at $r_{\rm min}\simeq\SI{11.66}{\angstrom}$ corresponds to the minimum of the effective radial confinement potential discussed below.
Although the dominant component of the confinement is radial, the annulus exhibits residual angular variations arising from the atomistic structure of the Ar-preplated interface.

\begin{figure}[t]
    \centering
    \includegraphics[width=\columnwidth]{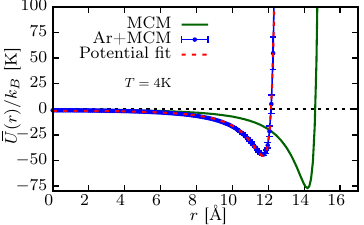}
    \caption{Radial helium confinement potential at $T=\SI{4}{\kelvin}$ for bare MCM-41 and Ar-preplated MCM-41. The Ar preplating shifts the attractive minimum toward the pore interior and reduces its depth relative to the bare pore. The red dashed curve shows the continuum cylindrical fit to the Ar-preplated potential described by Eq.~\eqref{eq:tpi_fit}.} \label{fig:tpi_radial}
\end{figure}

In order to isolate the dominant radial component of the confinement, we additionally average over the angular coordinate,
\begin{equation}
    \,\overline{\!U}(r)=\frac{1}{N_\theta N_z}\sum_{\theta,z}\langle U(r,\theta,z)\rangle,
\end{equation}
where $N_\theta$ is the number of sampled angular points at each radial
position. Figure~\ref{fig:tpi_radial} compares the resulting radial confinement potentials for bare and Ar-preplated MCM-41. For the bare pore, the attractive minimum lies near the silica surface at $r_{\rm min}^{\rm MCM}\simeq\SI{14.17}{\angstrom}$. Upon Ar preplating, the minimum shifts toward the pore interior to $r_{\rm min}^{\rm Ar+MCM}\simeq\SI{11.66}{\angstrom}$ and its depth is reduced to approximately $\overline{U}(r_{\rm min})/k_B=\SI{-44.16}{\kelvin}$. Thus, the Ar monolayer modifies both the spatial location and the strength of the helium adsorption potential. Away from the wall, the confinement becomes considerably flatter toward the pore center.

The radial confinement potential $\,\overline{\!U}(r)/k_B$ of the Ar-preplated pore can be fit to an analytic continuum model for a helium atom inside a perfect cylindrical cavity of radius $R$ carved inside an infinite continuous medium of density $n$ \cite{Zhang:2004,Nichols:2020},
\begin{equation}
U_{\rm eff}(r)=\frac{\pi n\varepsilon\sigma^3}{3}\left[\left(\frac{\sigma}{R}\right)^9\!\!u_9\!\left(\frac{r}{R}\right)-\left(\frac{\sigma}{R}\right)^3\!\!u_3\!\left(\frac{r}{R}\right)\right]+U_{0},
\label{eq:tpi_fit}
\end{equation}
where $\sigma$ is the hard-core distance, $\varepsilon$ controls the overall interaction strength, and $U_{0}$ is a constant energy offset. $u_3(x)$ and $u_9(x)$ are dimensionless geometric shape functions of the continuum cylinder model~\cite{Nichols:2020}. We observe that this effective model provides an excellent description of the averaged potential data shown in Fig.~\ref{fig:tpi_radial}, reproducing both the position and the depth of the radial minimum with high accuracy. The parameters determined from the fit are shown in Table~\ref{tab:TPI_fit}. This demonstrates that, although the underlying configuration of argon-preplated MCM-41 retains atomistic corrugation (as seen in Fig.~\ref{fig:tpi_heatmap}), the average helium confinement can be represented by a smooth effective radial potential. 
\begin{table}[h]
    \renewcommand{\arraystretch}{1.5}
    \setlength\tabcolsep{8pt}
    \begin{tabular}{@{}llll@{}} 
        \toprule
        $\sigma\; \qty[\si{\angstrom}]$ & $R\; \qty[\si{\angstrom}]$ & $n\varepsilon/\kB\; \qty[\si{\kelvin\angstrom^{-3}}]$ & $U_{0}/\kB\; \qty[\si{\kelvin}]$\\
        \midrule 
        $3.01 \pm 0.27$  & $14.28\pm 0.19$ & $0.64\pm 0.21$ & $0.14\pm 0.65$\\
        \bottomrule
    \end{tabular}
    \caption{\label{tab:TPI_fit}The fitting parameters for the continuum cylindrical model of the confinement potential for MCM-41 preplated with argon gas described by Eq.~\eqref{eq:tpi_fit}. Parameters were determined by fitting to the radially averaged test-particle insertion results shown in Fig.~\ref{fig:tpi_radial}.}
\end{table}

While the radial average captures the dominant confinement, the transverse potential shows that microscopic corrugation remains. To determine how Ar preplating modifies these fluctuations, we compare the corrugation of the bare and Ar-preplated confinement potentials relative to their respective adsorption minima. Since preplating shifts the radial minimum, we introduce the relative coordinate $\Delta r = r - r_{\rm min}$, such that $\Delta r=0$ corresponds to the minimum of each radial potential.
\begin{figure}[t]
    \centering
    \includegraphics[width=\columnwidth]{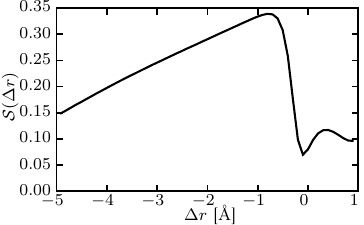}
    \caption{Screening factor $\mathcal{S}(\Delta r)$ is obtained from the RMS corrugation of the helium confinement potential in bare and Ar-preplated MCM-41. The radial coordinate is measured relative to the minimum of the corresponding radial potential, $\Delta r=r-r_{\rm min}$. Positive values indicate suppression of the corrugation upon Ar preplating. The maximum screening of $\simeq 34\%$ is observed at $\Delta r\simeq -\SI{0.8}{\angstrom}$.}\label{fig:screening_factor}
\end{figure}

At fixed radial position, we quantify the microscopic corrugation by the
root-mean-square (RMS) variation of the potential over the angular and axial coordinates, averaged over the 10 configurations:
\begin{equation} 
\Delta U_{\rm RMS}(r) = \left\langle \sqrt{\displaystyle\frac{1}{N_\theta N_z}\sum_{\theta,z}\left[ U_{\alpha}(r,\theta,z) - \overline{U}_{\alpha}(r) \right]^2}\right\rangle, \label{eq:rms_corrugation}
\end{equation}
where $\alpha$ represents a frozen Ar-preplated host realization. For the Ar-preplated pore, the RMS corrugation is first averaged over the two snapshots associated with each seed and subsequently over the five independent seeds. For bare MCM-41, the RMS corrugation is obtained using the same definition for the corresponding fixed host configuration. The reduction in corrugation upon preplating is quantified through the screening factor defined as
\begin{equation}
    \mathcal{S}(\Delta r) = 1-\frac{\Delta U_{\rm RMS}^{\rm Ar+MCM}(\Delta r)}{\Delta U_{\rm RMS}^{\rm MCM}(\Delta r)}. 
\end{equation}
Here, $\mathcal{S}>0$ indicates suppression of the microscopic corrugation by the Ar layer, whereas $\mathcal{S}=0$ corresponds to no change. As shown in Fig.~\ref{fig:screening_factor}, the screening depends on position relative to the potential minimum and reaches a maximum of approximately $34\%$ near $\Delta r\simeq-\SI{0.8}{\angstrom}$.  

\begin{figure}[t]
    \centering
    \includegraphics[width=\columnwidth]{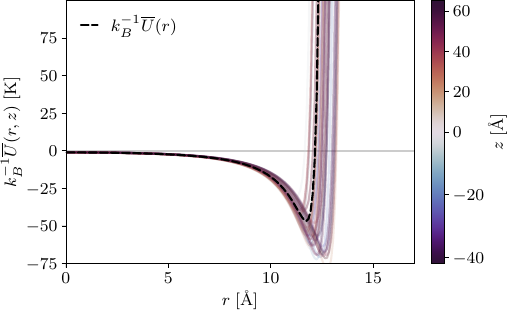}
    \caption{Azimuthally averaged helium confinement potential $\,\overline{\!U}(r,z)$ plotted as a function of radial position for different values of the axial coordinate $z$. The color scale denotes the axial position. The black dashed curve is the fully averaged radial potential $\,\overline{\!U}(r)$ as shown in Fig.~\ref{fig:tpi_radial}.}
    \label{fig:z_disorder}
\end{figure}

Although Ar preplating reduces the overall corrugation of the confinement potential,  variations remain along the pore axis. To characterize this axial dependence, we evaluated the azimuthally averaged potential
\begin{equation}
    \,\overline{\!U}(r,z) = \frac{1}{N_\theta} \sum_{\theta}\langle U(r,\theta,z)\rangle.
\end{equation}
Figure~\ref{fig:z_disorder} shows $\,\overline{\!U}(r,z)$ as a function of radial position for various positions along the axial direction. The curves remain nearly coincident toward the pore center but show appreciable variations near the attractive annular region. In particular, both the depth and shape of the radial minimum depend on the axial position, depicting that the confinement is not perfectly translationally invariant along the pore axis.

\subsection{The Corrugated Confinement Potential Energy Surface}
\begin{figure}[h]
    \centering
    \includegraphics[width=\columnwidth]{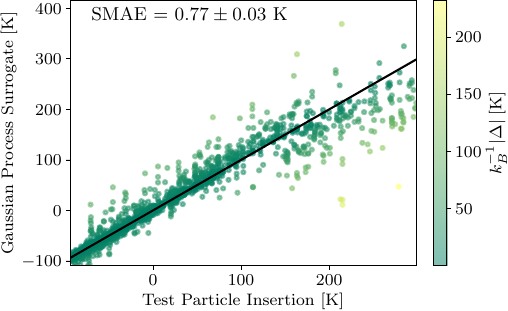}
    \caption{Predicted versus actual energies for the Gaussian Process surrogate model for helium confined inside argon preplated MCM-41. The reported uncertainty in the sample mean average error is the standard error across $M = 8$ realizations.}
    \label{fig:gpparity}
\end{figure}

To further quantify how well the effective smooth Lennard-Jones potential in Eq.~\eqref{eq:tpi_fit} approximates the corrugated potential of Fig.~\ref{fig:tpi_heatmap} we compute the difference $\delta U(r,\theta) = \,\overline{\!U}(r,\theta) - U_{\rm eff}(r)$ and probe its complexity by expanding in terms of a Fourier series, 
\begin{equation}
    \delta U(r,\theta) = \sum_{n=1}^{n_{\rm max}} \qty[A_n(r)\cos(n\theta) + B_n(r)\sin(n\theta)]\, . 
\label{eq:deltaU}
\end{equation}
The maximum Fourier mode $n_{\rm max}$ depends on the density of sampled points with a fixed radius and is given by the Nyquist frequency. 
\begin{figure}[h]
    \centering
    \includegraphics[width=\columnwidth]{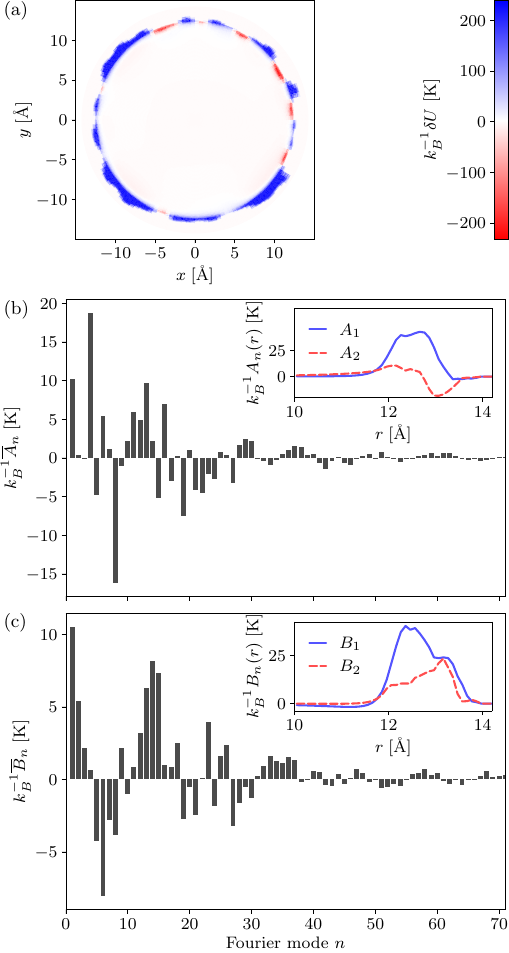}
    \caption{The disorder $\delta U$ and the average contribution (Eq.~\eqref{eq:meanAB}) of each Fourier coefficient to the disorder. The insets in panels (b) and (c) show the radial profile of the first and second modes in the expansion highlighting the role of the Ar and MCM-41 pore in surface corrugation.}
    \label{fig:disorder}
\end{figure}
From Fig.~\ref{fig:disorder} it is clear the corrugation is highly anisotropic, with both the sign and magnitude of deviations from the radially averaged potential having a strong angular dependence as shown in panel (a).  This can be captured by the radially averaged Fourier coefficients
\begin{equation}
    \overline{X}_n = \frac{1}{R_\text{pore} - 10} \int_{10}^{R_\text{pore}} dr X_n, \qquad X = A,B
\label{eq:meanAB}
\end{equation}
plotted in panel $(b)$ and $(c)$ which show oscillating non-trivial contributions, even up to very large orders. The absence of any outlying peaks and a broad spectrum for the average contribution indicate the presence of disorder at several length scales. The insets show the first and second ($n=1,2$) coefficients and their structure,  demonstrating that the corrugation arises primarily near the wall. Such a function is notoriously difficult to model, with features at multiple length scales motivating that any GP regression procedure will require a kernel able to capture the combination of short and long range disorder. As a result, we choose a kernel composed of a linear combination of two Mat{\'e}rn kernels $K_\nu$ 
\begin{equation}
    \vb{k}(\vb*{r},\vb*{r'}) = K_{5/2}(\vb*{r},\vb*{r'}) + K_{3/2}(\vb*{r},\vb*{r'})
\end{equation}
where 
\begin{multline}
    K_{\nu}(\vb*{r},\vb*{r'}) = \frac{A_{\nu}}{\Gamma(\nu)2^{\nu - 1}}\left(\frac{\sqrt{2\nu}}{\ell_{\nu}}|\vb*{r} - \vb*{r'}|\right)^{\nu} \\
    \times J_{\nu}\left(\frac{\sqrt{2\nu}}{\ell_{\nu}}|\vb*{r} - \vb*{r'}|\right)
\end{multline}
and $\ell_{\nu}$, $A_{\nu}$  are kernel hyperparameters to be learned while $J_\nu$ is the modified Bessel function of the second kind. The different values of $\nu$ affect the smoothness (differentiability) of the functions and each of them capture the corrugation at different length scales creating an accurate surrogate for the confinement potential. Training the surrogate on 80\% of our TPI data, we test the accuracy of reproduction by computing the sample mean average error (SMAE) \cite{Akram2026}, 
\begin{equation}
    \text{SMAE} = \frac{1}{M\, n_{\rm test}} \sum_{\alpha=1}^{M}\sum_{i=1}^{n_{\rm test}} \abs{{U}^{(\alpha)}(\vb*{r}_{\alpha_i}) - U_{\alpha_i}}\, ,
\end{equation}
where $\alpha_i$ is an integer index corresponding to a point in the test set where we have ground truth data, $U^{(\alpha)}$ is the GP model at realization $\alpha$ and $M$ is the number of different random train/test splits of the data. The resulting parity plot across random splits of the dataset is shown in Fig.~\ref{fig:gpparity}.
Points falling on the diagonal represent perfect reconstruction via the GP model and a small SMAE of $\SI{0.77}{\kelvin}$ implies that the surrogate can be used to accurately describe the corrugation found in fixed $z$-slices of Ar preplated MCM-41 nanopores. Extending to a surrogate for the full $z$-dependnece is hindered by the size of the full TPI dataset due to the matrix inverse in the Gaussian Process reconstruction (Eq.~\eqref{Eq:posteriormean}). 

% ==============================================================================

\section{Discussion}
\label{sec:discussion}

The central result of this work is an experimentally constrained connection between the preparation of an Ar-preplated MCM-41 pore and the microscopic one-body confinement potential experienced by a helium atom within it. The agreement of the calculated adsorption isotherm with experiment, together with the reproduction of the dominant features of the measured static structure factor, indicates that the atomistic model considered here captures the essential energetics and local structure of the adsorbed Ar monolayer. The remaining differences in the finer structure of $S(q)$ also emphasize the sensitivity of the adsorption environment to microscopic surface structure and to forms of sample heterogeneity that are absent from the single MCM-41 structure used here.

We find that the Ar monolayer does more than simply reduce the available pore radius. Test-particle insertion shows that it substantially screens the strongly corrugated silica surface and transfers the helium adsorption minimum to an annular region within the pore. After averaging over the angular and axial directions, the resulting confinement is accurately described by a smooth cylindrical potential of the form used in earlier simulations of Ar-preplated MCM-41 \cite{Nichols:2020}. This provides microscopic support for the effective radial description and identifies its parameters directly from an atomistic model. At the same time, the quality of the radially averaged fit should not be interpreted as evidence that the microscopic potential is featureless. The Fourier analysis and Gaussian process representation show that residual corrugation persists over several spatial scales and cannot be represented by a small number of long-wavelength angular modes.

This is important for confinement of helium at low temperature as the minimum of the one-body potential does not, by itself, determine the equilibrium helium density profile or its effective dimensionality. Helium-helium interactions and zero-point motion will determine how the annular adsorption region is populated, whether an additional liquid core develops near the pore axis, and how strongly the different regions are coupled by particle exchange. Residual corrugation may modify the local density and compressibility, pin defects in the adsorbed layer \cite{DelMaestro:2026fo}, or reduce exchanges between the annular shell and the central fluid. In a quasi-one-dimensional regime, these microscopic changes can renormalize the sound velocity and interactions of the emergent quantum hydrodynamics and thereby affect both the range over which behavior can be observed and the crossover to more strongly localized behavior \cite{DelMaestro:2022}. A description based only on an effective pore radius and adsorption depth may therefore capture the dominant geometry while missing disorder that is relevant to the many-body correlations.

Thus, the microscopic framework presented here provides a realistic starting point for future quantum many-body simulations. In particular, the Gaussian process representation of the full potential offers a path towards a systematic study of the effects of corrugation on a possible emergent one-dimensional quantum liquid of helium inside preplated nanoporous materials.

\section{Data and Code Availability}
All code and data \cite{paperrepo} needed to reproduce the results of this study are available online.

% ==============================================================================

\section{Acknowledgments}
The authors thank Ioannis Sgouralis for helpful scientific discussions.
This work was supported by the U.S. Department of Energy, Office of Science, Office of Basic Energy Sciences, under Award Number DE-SC0024333.  N.S.N.  was supported by the Office of Science, U.S. Department of Energy, under contract DE-AC02-06CH11357.

\appendix
\section{Monolayer identification}\label{ssec:monolayer_identification}
We identify the first and second adsorption layers from the radial density
profile of Ar inside the pore. The outermost density peak, adjacent to the
silica wall, is assigned to the first layer, while the next peak at smaller
radius is assigned to the second layer. The local minimum separating these
peaks defines the radial boundary used to calculate their occupancies,
$N_1(\mu)$ and $N_2(\mu)$. 
\begin{figure}[h]
    \centering
    \includegraphics[width=\columnwidth]{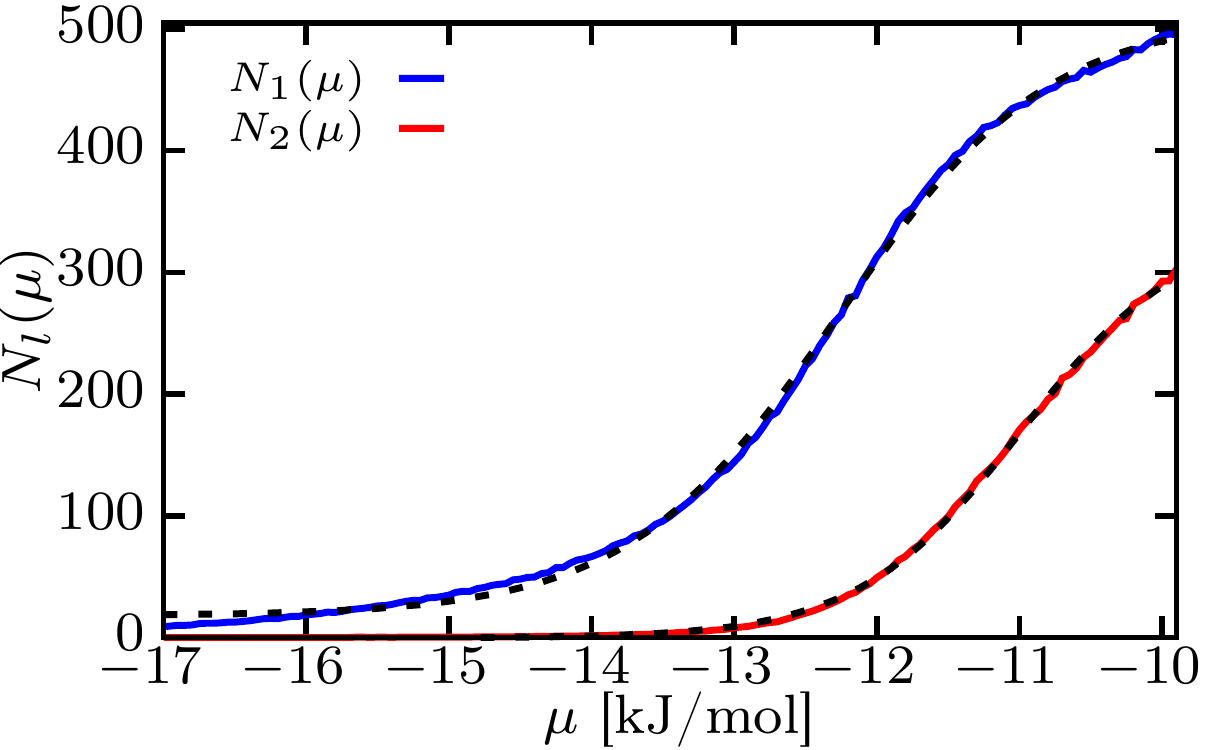}
    \caption{\label{fig:N1_N2_vs_mu_with_fits} Occupancies of the first and second adsorption layers, $N_1(\mu)$ and $N_2(\mu)$, as functions of chemical potential. The solid curves show the occupancies obtained from radial densities (Fig.~\ref{fig:radial_density}), and the black dashed curves show the corresponding sigmoidal fits.}
\end{figure}

As demonstrated in Fig.~\ref{fig:N1_N2_vs_mu_with_fits}, the adsorption initially
occurs predominantly in the first layer, with $N_1(\mu)$ increasing while $N_2(\mu)$ remains small. The two layers, however, do not fill independently or in a strictly sequential manner. The second-layer occupancy begins to increase before the first layer has reached saturation, producing a range of chemical potentials over which both layers grow simultaneously. Thus, monolayer completion cannot be identified as the point at which $N_1(\mu)$ becomes constant.

To distinguish the regimes in which first or second layer growth dominates, we fit both occupancy curves with smooth sigmoidal functions and calculate their derivatives with respect to chemical potential.
\begin{figure}[h]
    \centering
    \includegraphics[width=\columnwidth]{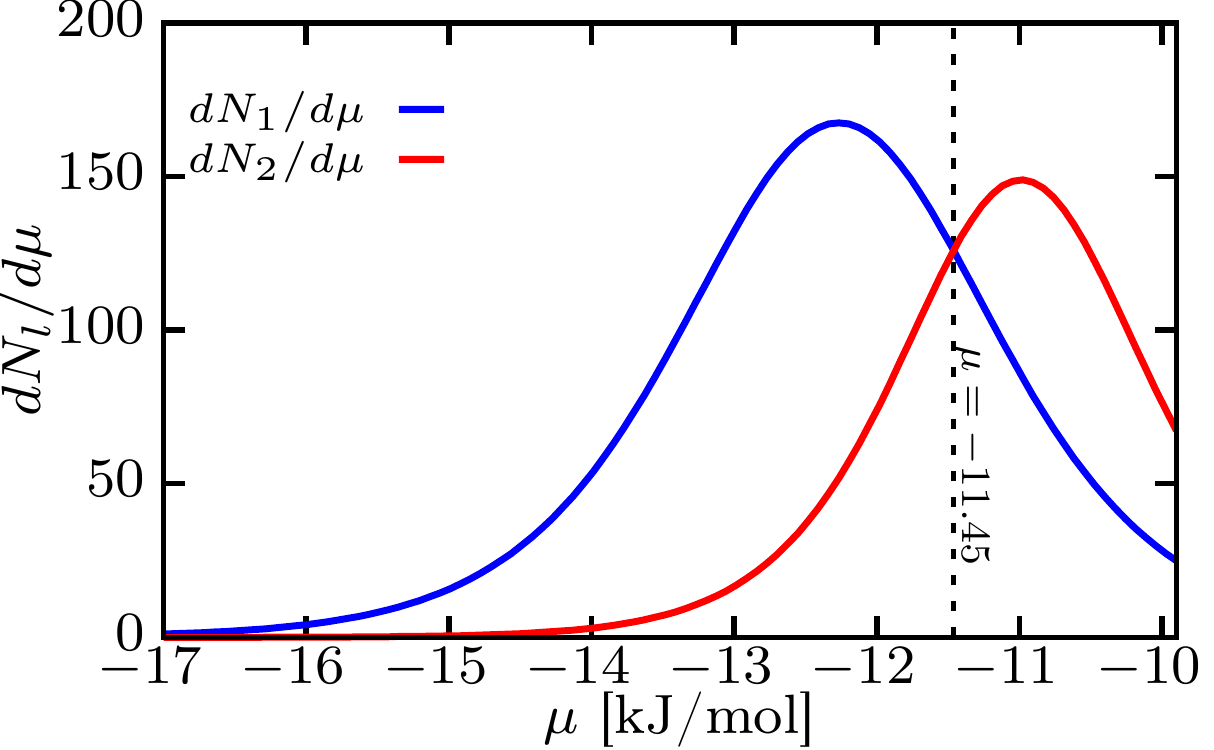}
    \caption{\label{fig:dN1_dN2_vs_mu} Derivatives of the first and second layer occupancies, \(dN_1/d\mu\) and \(dN_2/d\mu\), obtained from the smooth sigmoidal fits shown in Fig.~\ref{fig:N1_N2_vs_mu_with_fits}. The vertical dashed line marks $\mu=\SI{-11.45}{\kilo\joule\per\mol}$, which we identify as the monolayer-completion point where the growth of the first layer begins to diminish while the second layer starts to increase rapidly.}
\end{figure}

Figure~\ref{fig:dN1_dN2_vs_mu} shows that $dN_1/d\mu$ reaches its maximum first and then decreases, while $dN_2/d\mu$ continues to increase. We define the monolayer-completion chemical potential by the crossing condition
\begin{equation} 
\left. \frac{dN_1}{d\mu} \right|_{\mu=\mu_{\rm mono}} = -\left. \frac{dN_2}{d\mu} \right|_{\mu=\mu_{\rm mono}} . 
\end{equation}
This criterion gives $\mu_{\rm mono}=\SI{-11.45}{\kilo\joule\per\mol}$, the value used for the monolayer configurations analyzed in the main text.

% ------------------------------------------------------------------------------
% ==============================================================================

\FloatBarrier

% \nocite{apsrev41Control}
% \bibliographystyle{apsrev4-2}
\bibliography{refs}

@article{Kresge:1992kd,
  author = {Kresge, C T and Leonowicz, M E and Roth, W J and Vartuli, J C and Beck, J S},
  title = {{Ordered mesoporous molecular sieves synthesized by a liquid-crystal template mechanism}},
  journal = {Nature},
  year = {1992},
  volume = {359},
  number = {6397},
  pages = {710--712},
  month = oct,
  doi = {10.1038/359710a0},
  url = {http://www.nature.com/doifinder/10.1038/359710a0},
}

@article{Andrea:2022,
author = {Carta, Paola and Cara, Claudio and Cannas, Carla and Scorciapino, Mariano Andrea},
title = {Experiments-Guided Modeling of MCM-41: Impact of Pore Symmetry on Gas Adsorption},
journal = {Advanced Materials Interfaces},
volume = {9},
number = {34},
pages = {2201591},
doi = {10.1002/admi.202201591},
url = {https://advanced.onlinelibrary.wiley.com/doi/abs/10.1002/admi.202201591},
year = {2022}
}

@article{DelMaestro:2026fo,
  title = {Friedel Oscillations in Nanoconfined $^{4}\mathrm{He}$},
  author = {Rosenow, Bernd and Del Maestro, Adrian},
  journal = {Phys. Rev. Lett.},
  volume = {136},
  issue = {14},
  pages = {146002},
  numpages = {7},
  year = {2026},
  month = {Apr},
  publisher = {American Physical Society},
  doi = {10.1103/flnv-3ts2},
  url = {https://link.aps.org/doi/10.1103/flnv-3ts2}
}

@article{Wada:2001jb,
  author = {Wada, Nobuo and Taniguchi, Junko and Ikegami, Hiroki and Inagaki, Shinji and Fukushima, Yoshiaki},
  title = {{Helium-4 Bose Fluids Formed in One-Dimensional 18 {\AA} Diameter Pores}},
  journal = {Phys. Rev. Lett.},
  year = {2001},
  volume = {86},
  number = {19},
  pages = {4322--4325},
  month = may,
  doi = {10.1103/PhysRevLett.86.4322},
  url = {http://link.aps.org/doi/10.1103/PhysRevLett.86.4322},
}

@article{Wada:2005uc,
  author = {Wada, Nobuo and Taniguchi, Junko and Matsushita, Taku and Toda, Ryo and Matsushita, Yuki and Ikegami, Hiroki and Hieda, Mitsunori and Yamaguchi, Akira and Ishimoto, Hidehiko},
  doi = {10.1016/j.jpcs.2005.05.069},
  journal = {J. Phys. Chem. Solids},
  number = {8-9},
  pages = {1512},
  publisher = {Elsevier {BV}},
  title = {{Z}ero- and one-dimensional $^4${H}e {B}ose fluids realized in nanometer pores},
  url = {https://linkinghub.elsevier.com/retrieve/pii/S0022369705001411},
  volume = {66},
  year = {2005}
}

@article{Ikegami:2005ec,
  author = {Ikegami, H. and Yamato, Y. and Okuno, T. and Taniguchi, J. and Wada, N.},
  doi = {10.1007/s10909-005-1546-2},
  journal = {J. Low Temp. Phys.},
  number = {1-2},
  pages = {171},
  publisher = {Springer Science and Business Media {LLC}},
  title = {{O}bservation of $^4${H}e {S}uperfluidity in 1.8 nm-{P}ores},
  url = {https://link.springer.com/article/10.1007%2Fs10909-005-1546-2},
  volume = {138},
  year = {2005}
}

@article{Toda:2007cv,
  author = {Toda, Ryo and Hieda, Mitsunori and Matsushita, Taku and Wada, Nobuo and Taniguchi, Junko and Ikegami, Hiroki and Inagaki, Shinji and Fukushima, Yoshiaki},
  title = {{Superfluidity of $^4${H}e in One and Three Dimensions Realized in Nanopores}},
  journal = {Phys. Rev. Lett.},
  year = {2007},
  volume = {99},
  number = {25},
  pages = {255301},
  month = dec,
  doi = {10.1103/PhysRevLett.99.255301},
  url = {https://doi.org/10.1103/PhysRevLett.99.255301},
}

@article{Taniguchi:2011bx,
  author = {Taniguchi, Junko and Fujii, Rina and Suzuki, Masaru},
  title = {{Superfluidity and BEC of liquid $^{4}$He confined in a nanometer-size channel}},
  journal = {Phys. Rev. B},
  year = {2011},
  volume = {84},
  number = {13},
  pages = {134511},
  month = oct,
  doi = {10.1103/PhysRevB.84.134511},
  url = {http://link.aps.org/doi/10.1103/PhysRevB.84.134511},
}

@article{Taniguchi:2013us,
  author = {Taniguchi, J and Demura, K and Suzuki, M},
  title = {{Dynamical superfluid response of $^{4}$He confined in a nanometer-size channel}},
  journal = {Phys. Rev. B},
  year = {2013},
  volume = {88},
  number = {1},
  pages = {014502},
  doi = {10.1103/PhysRevB.88.014502},
  url = {https://doi.org/10.1103/PhysRevB.88.014502}
}

@article{Yager:2013cv,
  author = {Yager, B and Ny{\'e}ki, J and Casey, A and Cowan, B P and Lusher, C P and Saunders, J},
  title = {{NMR Signature of One-Dimensional Behavior of $^{3}$He in Nanopores}},
  journal = {Phys. Rev. Lett.},
  year = {2013},
  volume = {111},
  number = {21},
  pages = {215303},
  month = nov,
  doi = {10.1103/PhysRevLett.111.215303},
  url = {http://link.aps.org/doi/10.1103/PhysRevLett.111.215303},
}

@article{Demura:2015hq,
  author = {Demura, Kenta and Taniguchi, Junko and Suzuki, Masaru},
  title = {{Dynamical Superfluid Response of $^3${H}e{\textendash}$^4${H}e Solutions Confined in a Nanometer-Size Channel}},
  journal = {J. Phys. Soc. Jpn.},
  year = {2015},
  volume = {84},
  number = {9},
  pages = {094604},
  month = sep,
  doi = {10.7566/JPSJ.84.094604},
  url = {https://doi.org/10.7566/JPSJ.84.094604},
}

@article{Demura:2017gy,
  author = {Demura, Kenta and Taniguchi, Junko and Suzuki, Masaru},
  title = {{Twofold Torsional Oscillator Experiments from Film to Pressurized Liquid $^4${H}e in a Nanometer-Size Channel}},
  journal = {J. Phys. Soc. Jap.},
  year = {2017},
  volume = {86},
  number = {11},
  pages = {114601},
  month = nov,
  publisher = {The Physical Society of Japan},
  doi = {10.7566/JPSJ.86.114601},
  url = {https://doi.org/10.7566/JPSJ.86.114601},
}

@article{Taniguchi:2018ip,
  doi = {10.1088/1742-6596/969/1/012005},
  url = {https://doi.org/10.1088/1742-6596/969/1/012005},
  year = {2018},
  publisher = {{IOP} Publishing},
  volume = {969},
  pages = {012005},
  author = {Kento Taniguchi and Junko Taniguchi and Masaru Suzuki},
  title = {{Torsional Oscillator Measurements of Liquid $^4${H}e Confined in 2.5-nm Channel of {FSM}}},
  journal = {J. Phys.: Conf. Ser.}
}

@article{Prisk:2013cu,
  author = {Prisk, Timothy R and Das, Narayan C and Diallo, Souleymane O and Ehlers, Georg and Podlesnyak, Andrey A and Wada, Nobuo and Inagaki, Shinji and Sokol, Paul E},
  title = {{Phases of superfluid helium in smooth cylindrical pores}},
  journal = {Phys. Rev. B},
  year = {2013},
  volume = {88},
  number = {1},
  pages = {014521},
  month = jul,
  doi = {10.1103/PhysRevB.88.014521},
  url = {http://link.aps.org/doi/10.1103/PhysRevB.88.014521},
}

@article{Bryan:2017hb,
  author = {Bryan, M S and Prisk, T R and Sherline, T E and Diallo, S O and Sokol, P E},
  title = {{Bulklike excitations in nanoconfined liquid helium}},
  journal = {Phys. Rev. B},
  year = {2017},
  volume = {95},
  number = {14},
  pages = {144509},
  month = apr,
  publisher = {American Physical Society},
  doi = {10.1103/PhysRevB.95.144509},
  url = {http://link.aps.org/doi/10.1103/PhysRevB.95.144509},
}

@article{Bryan:2018vb,
  title = {Maxon and roton measurements in nanoconfined $^{4}\mathrm{He}$},
  author = {Bryan, M. S. and Sokol, P. E.},
  journal = {Phys. Rev. B},
  volume = {97},
  issue = {18},
  pages = {184511},
  numpages = {8},
  year = {2018},
  month = {May},
  publisher = {American Physical Society},
  doi = {10.1103/PhysRevB.97.184511},
  url = {https://link.aps.org/doi/10.1103/PhysRevB.97.184511}
}

@article{Bossy:2019qd,
  author = {Bossy, Jacques and Ollivier, Jacques and Glyde, H.~R.},
  doi = {10.1103/physrevb.99.165425},
  journal = {Phys. Rev. B},
  number = {16},
  pages = {165425},
  publisher = {American Physical Society ({APS})},
  title = {{{P}honons, rotons, and localized {B}ose-{E}instein condensation in liquid $^4${H}e
 confined in nanoporous {F{S}M}-16}},
  url = {https://journals.aps.org/prb/abstract/10.1103/PhysRevB.99.165425},
  volume = {99},
  year = {2019}
}

@article{Taniguchi:2020ln,
  author = {Taniguchi, Junko and Taniguchi, Kento and Kanno, Kousuke and Suzuki, Masaru},
  doi = {10.1007/s10909-020-02355-z},
  journal = {J. Low Temp. Phys.},
  pages = {139},
  publisher = {Springer Science and Business Media {LLC}},
  title = {{{P}ossible {T}hermodynamical {P}hase {S}lips in {S}uperfluid $^4${H}e {C}onfined in a 2.5-nm {C}hannel of {F{S}M}}},
  url = {https://doi.org/10.1007/s10909-020-02355-z},
  volume = {201},
  year = {2020}
}

@article{Huan:2020ya,
  author = {Huan, C. and Adams, J. and Lewkowitz, M. and Masuhara, N. and Candela, D. and Sullivan, N.~S.},
  doi = {10.1007/s10909-020-02358-w},
  journal = {J. Low Temp. Phys.},
  pages = {146},
  year = {2020},
  publisher = {Springer Science and Business Media {LLC}},
  title = {{N{M}R} {S}tudies of the {D}ynamics of 1{D} $^3${H}e in $^4${H}e {P}lated {MCM}-41},
  url = {https://link.springer.com/article/10.1007%2Fs10909-020-02358-w},
  volume = {201},
}

@article{paperrepo,
	journal = {{Github Repository}},
	title = {{Interface Engineering of Helium Confinement in Argon-Preplated MCM-41 Nanopores}},
	author = {Rahul Soni and Adrian {Del Maestro}},
	url = {https://github.com/DelMaestroGroup/papers-code-ArMCMTPI},
	year = {2026},
	doi = {10.5281/zenodo.21814518}
}

@article{DelMaestro:2011,
  title = {$^{4}\mathrm{He}$ Luttinger Liquid in Nanopores},
  author = {Del Maestro, Adrian and Boninsegni, Massimo and Affleck, Ian},
  journal = {Phys. Rev. Lett.},
  volume = {106},
  issue = {10},
  pages = {105303},
  numpages = {4},
  year = {2011},
  month = {Mar},
  publisher = {American Physical Society},
  doi = {10.1103/PhysRevLett.106.105303},
  url = {https://link.aps.org/doi/10.1103/PhysRevLett.106.105303}
}

@article{DelMaestro:2022,
	author = {Del Maestro, Adrian and Nichols, Nathan S. and Prisk, Timothy R. and Warren, Garfield and Sokol, Paul E.},
	da = {2022/06/07},
	doi = {10.1038/s41467-022-30752-3},
	id = {Del Maestro2022},
	isbn = {2041-1723},
	journal = {Nat. Commun.},
	number = {1},
	pages = {3168},
	title = {Experimental realization of one dimensional helium},
	url = {https://doi.org/10.1038/s41467-022-30752-3},
	volume = {13},
	year = {2022}}

@article{Paul:2026,
  title = {Localization and wetting of $^{4}\mathrm{He}$ inside preplated nanopores},
  author = {Paul, Sutirtha and Lakoba, Taras and Sokol, Paul E. and Del Maestro, Adrian},
  journal = {Phys. Rev. B},
  volume = {113},
  issue = {7},
  pages = {075433},
  numpages = {10},
  year = {2026},
  month = {Feb},
  publisher = {American Physical Society},
  doi = {10.1103/bbm4-lz85},
  url = {https://link.aps.org/doi/10.1103/bbm4-lz85}
}

@article{Debye:1915,
  author  = {Debye, P.},
  title   = {Zerstreuung von R{\"o}ntgenstrahlen},
  journal = {Annalen der Physik},
  volume  = {351},
  number  = {6},
  pages   = {809--823},
  year    = {1915},
  doi     = {10.1002/andp.19153510606}
}

@article{plimpton:1995,
title = {Fast Parallel Algorithms for Short-Range Molecular Dynamics},
journal = {J. Comp. Phys.},
volume = {117},
number = {1},
pages = {1-19},
year = {1995},
issn = {0021-9991},
doi = {10.1006/jcph.1995.1039},
url = {https://www.sciencedirect.com/science/article/pii/S002199918571039X},
author = {Steve Plimpton}
}

@article{thompson:2022,
title = {LAMMPS - a flexible simulation tool for particle-based materials modeling at the atomic, meso, and continuum scales},
journal = {Comp. Phys. Comm.},
volume = {271},
pages = {108171},
year = {2022},
issn = {0010-4655},
doi = {10.1016/j.cpc.2021.108171},
url = {https://www.sciencedirect.com/science/article/pii/S0010465521002836},
author = {Aidan P. Thompson and H. Metin Aktulga and Richard Berger and Dan S. Bolintineanu and W. Michael Brown and Paul S. Crozier and Pieter J. {in 't Veld} and Axel Kohlmeyer and Stan G. Moore and Trung Dac Nguyen and Ray Shan and Mark J. Stevens and Julien Tranchida and Christian Trott and Steven J. Plimpton}
}

@article{furukawa:2005,
 author = {Furukawa, Shin-ichi and Nishiumi, Toshihiro and Aoyama, Naoki and Nitta, Tomoshige and Nakano, Masayoshi},
 doi = {10.1252/jcej.38.999},
 journal = {J. Chem. Eng. Jpn.},
 number = {12},
 title = {{A Molecular Simulation Study on Adsorption of Acetone/Water in Mesoporous Silicas Modified by Pore Surface Silylation}},
 pages = {999},
 url = {https://www.jstage.jst.go.jp/article/jcej/38/12/38_12_999/_article},
 volume = {38},
 year = {2005}
}

@article{yun:2002,
 author = {Yun, Jeong-Ho and D{\"u}ren, Tina and Keil, Frerich J. and Seaton, Nigel A.},
 doi = {10.1021/la0155855},
 journal = {Langmuir},
 number = {7},
 pages = {2693},
 publisher = {American Chemical Society (ACS)},
 title = {{A}dsorption of {M}ethane, {E}thane, and {T}heir {B}inary {M}ixtures on {M}CM-41: {E}xperimental {E}valuation of {M}ethods for the {P}rediction of {A}dsorption {E}quilibrium},
 url = {https://pubs.acs.org/doi/10.1021/la0155855},
 volume = {18},
 year = {2002}
}

@article{Nichols:2020,
  title = {Dimensional reduction of helium-4 inside argon-plated MCM-41 nanopores},
  author = {Nichols, Nathan S. and Prisk, Timothy R. and Warren, Garfield and Sokol, Paul and Del Maestro, Adrian},
  journal = {Phys. Rev. B},
  volume = {102},
  issue = {14},
  pages = {144505},
  numpages = {13},
  year = {2020},
  month = {Oct},
  publisher = {American Physical Society},
  doi = {10.1103/PhysRevB.102.144505},
  url = {https://doi.org/10.1103/PhysRevB.102.144505}
}

@article{Ugliengo:2008ks,
  author = {Ugliengo, P and Sodupe, M and Musso, F and Bush, I J and Orlando, R and Dovesi, R},
  title = {{Realistic Models of Hydroxylated Amorphous Silica Surfaces and MCM-41 Mesoporous Material Simulated by Large-scale Periodic B3LYP Calculations}},
  journal = {Adv. Mater.},
  year = {2008},
  volume = {20},
  number = {23},
  pages = {4579--4583},
  month = dec,
  publisher = {Wiley-Blackwell},
  doi = {10.1002/adma.200801489},
  url = {http://doi.wiley.com/10.1002/adma.200801489},
}

@book{1964:HirschfelderBook,
  title={Molecular theory of gases and liquids},
  author={Hirschfelder, Joseph O and Curtiss, Charles F and Bird, Robert Byron and Mayer, Maria Goeppert},
  year={1964},
  publisher={Wiley New York}
}

@article{mcm41SA:2008,
    url = {https://www.sigmaaldrich.com/technical-documents/articles/material-matters/mesoporous-materials.html},
    author = {Sigma-Aldrich},
    journal = {Mater. Matters},
    title = {{Synthesis of Mesoporous Materials}},
    year = {2008},
    volume = {3.1},
    issn = {1558-366X},
    pages = {17},
}

@article{Brunauer:1938pz,
 author = {Brunauer, Stephen and Emmett, P.~H. and Teller, Edward},
 doi= {10.1021/ja01269a023},
 journal = {J. Am. Chem. Soc.},
 number = {2},
 pages = {309},
 publisher = {American Chemical Society ({ACS})},
 title = {{A}dsorption of {G}ases in {M}ultimolecular {L}ayers},
 url = {https://pubs.acs.org/doi/abs/10.1021/ja01269a023},
 volume = {60},
 year = {1938}
}

@article{Jaroniec:1999mi,
 author = {Jaroniec, Mietek and Kruk, Michal and Olivier, James P.},
 doi = {10.1021/la990136e},
 journal = {Langmuir},
 number = {16},
 pages = {5410},
 publisher = {American Chemical Society ({ACS})},
 title = {{S}tandard {N}itrogen {A}dsorption {D}ata for {C}haracterization of {N}anoporous {S}ilicas},
 url = {https://pubs.acs.org/doi/10.1021/la990136e},
 volume = {15},
 year = {1999}
}

@article{Copley:2003dc,
  doi = {10.1016/s0301-0104(03)00124-1},
  url = {https://doi.org/10.1016/s0301-0104(03)00124-1},
  year = {2003},
  month = aug,
  publisher = {Elsevier {BV}},
  volume = {292},
  number = {2-3},
  pages = {477--485},
  author = {J.R.D. Copley and J.C. Cook},
  title = {{The Disk Chopper Spectrometer at {NIST}: a new instrument for quasielastic neutron scattering studies}},
  journal = {Chem. Phys.}
}

@article{Azuah:2009cs,
 author = {Azuah, Richard Tumanjong and Kneller, Larry R. and Qiu, Yiming and Tregenna-Piggott, Philip L.~W. and Brown, Craig M. and Copley, John R.~D. and Dimeo, Robert M.},
 doi = {10.6028/jres.114.025},
 journal = {J. Res. Natl. Inst. Stand. Technol.},
 number = {6},
 pages = {341},
 publisher = {National Institute of Standards and Technology ({NIST})},
 title = {{D{A}VE}: {A} Comprehensive {S}oftware {S}uite for the {R}eduction, {V}isualization, and {A}nalysis of {L}ow {E}nergy {N}eutron {S}pectroscopic {D}ata},
 url = {https://dx.doi.org/10.6028/jres.114.025},
 volume = {114},
 year = {2009}
}

@book{Rasmussen:2005gp,
  title={Gaussian Processes for Machine Learning},
  author={Rasmussen, C.E. and Williams, C.K.I.},
  isbn={9780262182539},
  lccn={2005053433},
  url={https://books.google.com/books?id=Tr34DwAAQBAJ},
  year={2005},
  address = {Cambridge, MA},
  publisher={MIT Press}
}

@article{Kolb2017,
  author  = {Kolb, Brian and Marshall, Paul and Zhao, Bin and Jiang, Bin and Guo, Hua},
  title   = {Representing Global Reactive Potential Energy Surfaces Using Gaussian Processes},
  journal = {The Journal of Physical Chemistry A},
  year    = {2017},
  volume  = {121},
  number  = {13},
  pages   = {2552--2557},
  doi     = {10.1021/acs.jpca.7b01182}
}

@article{Uteva2018,
    author = {Uteva, Elena and Graham, Richard S. and Wilkinson, Richard D. and Wheatley, Richard J.},
    title = {Active learning in Gaussian process interpolation of potential energy surfaces},
    journal = {J. Chem. Phys.},
    volume = {149},
    number = {17},
    pages = {174114},
    year = {2018},
    month = {11},
    issn = {0021-9606},
    doi = {10.1063/1.5051772},
    url = {https://doi.org/10.1063/1.5051772},
}

@article{Wiens2019MFGP,
  author  = {Wiens, Avery E. and Copan, Andreas V. and Schaefer, Henry F.},
  title   = {Multi-fidelity Gaussian process modeling for chemical energy surfaces},
  journal = {Chemical Physics Letters: X},
  year    = {2019},
  volume  = {3},
  pages   = {100022},
  doi     = {10.1016/j.cpletx.2019.100022}
}

@article{Deringer2021ky,
 author = {Deringer, Volker L. and Bart{\'o}k, Albert P. and Bernstein, Noam
           and Wilkins, David M. and Ceriotti, Michele and Cs{\'a}nyi, G{\'a}bor},
 doi = {10.1021/acs.chemrev.1c00022},
 journal = {Chem. Rev.},
 number = {16},
 pages = {10073},
 publisher = {American Chemical Society (ACS)},
 title = {{G}aussian {P}rocess {R}egression for {M}aterials and {M}olecules},
 url = {https://pubs.acs.org/doi/10.1021/acs.chemrev.1c00022},
 volume = {121},
 year = {2021}
}

@article{Schneider2023,
    author = {Schneider, Moritz and Born, Daniel and Kästner, Johannes and Rauhut, Guntram},
    title = {Positioning of grid points for spanning potential energy surfaces—How much effort is really needed?},
    journal = {J. Chem. Phys.},
    volume = {158},
    number = {14},
    pages = {144118},
    year = {2023},
    month = {04},
    issn = {0021-9606},
    doi = {10.1063/5.0146020},
    url = {https://doi.org/10.1063/5.0146020},
}

@article{Akram2026,
    author = {Akram, Shahzad and Paul, Sutirtha and Kovacs, Collin and Maroulas, Vasileios and Del Maestro, Adrian and Vogiatzis, Konstantinos D.},
    title = {Accurate helium-benzene potential: From CCSD(T) to Gaussian process regression},
    journal = {The Journal of Chemical Physics},
    volume = {164},
    number = {11},
    pages = {114108},
    year = {2026},
    month = {03},
    issn = {0021-9606},
    doi = {10.1063/5.0322444},
}

@article{Carta:2023,
  author  = {Carta, Paola and Scorciapino, Mariano Andrea},
  title   = {Surface Heterogeneity Affects Adsorption Selectivity for {CO$_2$} over {CH$_4$} in Bare Mesostructured Silica with {2D} Hexagonal Symmetry and Different Pore Size},
  journal = {Advanced Materials Interfaces},
  year    = {2023},
  volume  = {10},
  number  = {26},
  pages   = {2300196},
  doi     = {10.1002/admi.202300196},
  url     = {https://doi.org/10.1002/admi.202300196}
}

@article{CychoszStruckhoff:2020,
  author  = {Cychosz Struckhoff, Katie and Thommes, Matthias and Sarkisov, Lev},
  title   = {On the Universality of Capillary Condensation and Adsorption Hysteresis Phenomena in Ordered and Crystalline Mesoporous Materials},
  journal = {Advanced Materials Interfaces},
  year    = {2020},
  volume  = {7},
  number  = {12},
  pages   = {2000184},
  doi     = {10.1002/admi.202000184},
  url     = {https://doi.org/10.1002/admi.202000184}
}

@article{Weinberger:2022,
  author  = {Weinberger, Christian and Zysk, Frederik and Hartmann, Marc and Kaliannan, Naveen K. and Keil, Waldemar and K{\"u}hne, Thomas D. and Tiemann, Michael},
  title   = {The Structure of Water in Silica Mesopores---Influence of the Pore Wall Polarity},
  journal = {Advanced Materials Interfaces},
  year    = {2022},
  volume  = {9},
  number  = {20},
  pages   = {2200245},
  doi     = {10.1002/admi.202200245},
  url     = {https://doi.org/10.1002/admi.202200245}
}

@article{Schlumberger:2021,
  author  = {Schlumberger, Carola and Thommes, Matthias},
  title   = {Characterization of Hierarchically Ordered Porous Materials by Physisorption and Mercury Porosimetry---A Tutorial Review},
  journal = {Advanced Materials Interfaces},
  year    = {2021},
  volume  = {8},
  number  = {4},
  pages   = {2002181},
  doi     = {10.1002/admi.202002181},
  url     = {https://doi.org/10.1002/admi.202002181}
}

@article{Deringer:2019AdvMater,
  author  = {Deringer, Volker L. and Caro, Miguel A. and Cs{\'a}nyi, G{\'a}bor},
  title   = {Machine Learning Interatomic Potentials as Emerging Tools for Materials Science},
  journal = {Advanced Materials},
  year    = {2019},
  volume  = {31},
  pages   = {1902765},
  doi     = {10.1002/adma.201902765},
  url     = {https://doi.org/10.1002/adma.201902765}
}

@article{Thommes:2015IUPAC,
  author  = {Thommes, Matthias and Kaneko, Katsumi and Neimark, Alexander V. and Olivier, James P. and Rodr{\'i}guez-Reinoso, Francisco and Rouquerol, Jean and Sing, Kenneth S. W.},
  title   = {Physisorption of Gases, with Special Reference to the Evaluation of Surface Area and Pore Size Distribution ({IUPAC} Technical Report)},
  journal = {Pure and Applied Chemistry},
  year    = {2015},
  volume  = {87},
  number  = {9--10},
  pages   = {1051--1069},
  doi     = {10.1515/pac-2014-1117},
  url     = {https://doi.org/10.1515/pac-2014-1117}
}

@article{Zhang:2004,
  author  = {Zhang, Xianren and Wang, Wenchuan and Jiang, Guangfeng},
  title   = {A Potential Model for Interaction between the {Lennard--Jones} Cylindrical Wall and Fluid Molecules},
  journal = {Fluid Phase Equilibria},
  year    = {2004},
  volume  = {218},
  number  = {2},
  pages   = {239--246},
  doi     = {10.1016/j.fluid.2004.01.005},
  url     = {https://doi.org/10.1016/j.fluid.2004.01.005}
}

@article{McNamara:2025,
  author  = {McNamara, Stephanie and Parajuli, Prabin and Paul, Sutirtha and Warren, Garfield and Del Maestro, Adrian and Sokol, Paul E.},
  title   = {Novel Experimental Platform to Realize One-Dimensional Quantum Fluids},
  journal = {Journal of Low Temperature Physics},
  year    = {2025},
  volume  = {221},
  pages   = {256--264},
  doi     = {10.1007/s10909-025-03329-9},
  url     = {https://doi.org/10.1007/s10909-025-03329-9}
}

%\includepdf[pages=-,offset=0 0]{supplement.pdf}

%\ifarXiv
    \foreach \x in {1,...,2}
    {
        \clearpage
        %\includepdf[pages={\x,{}}]{\supplementfilename.pdf}
        \includepdf[pages=\x, offset=0 0]{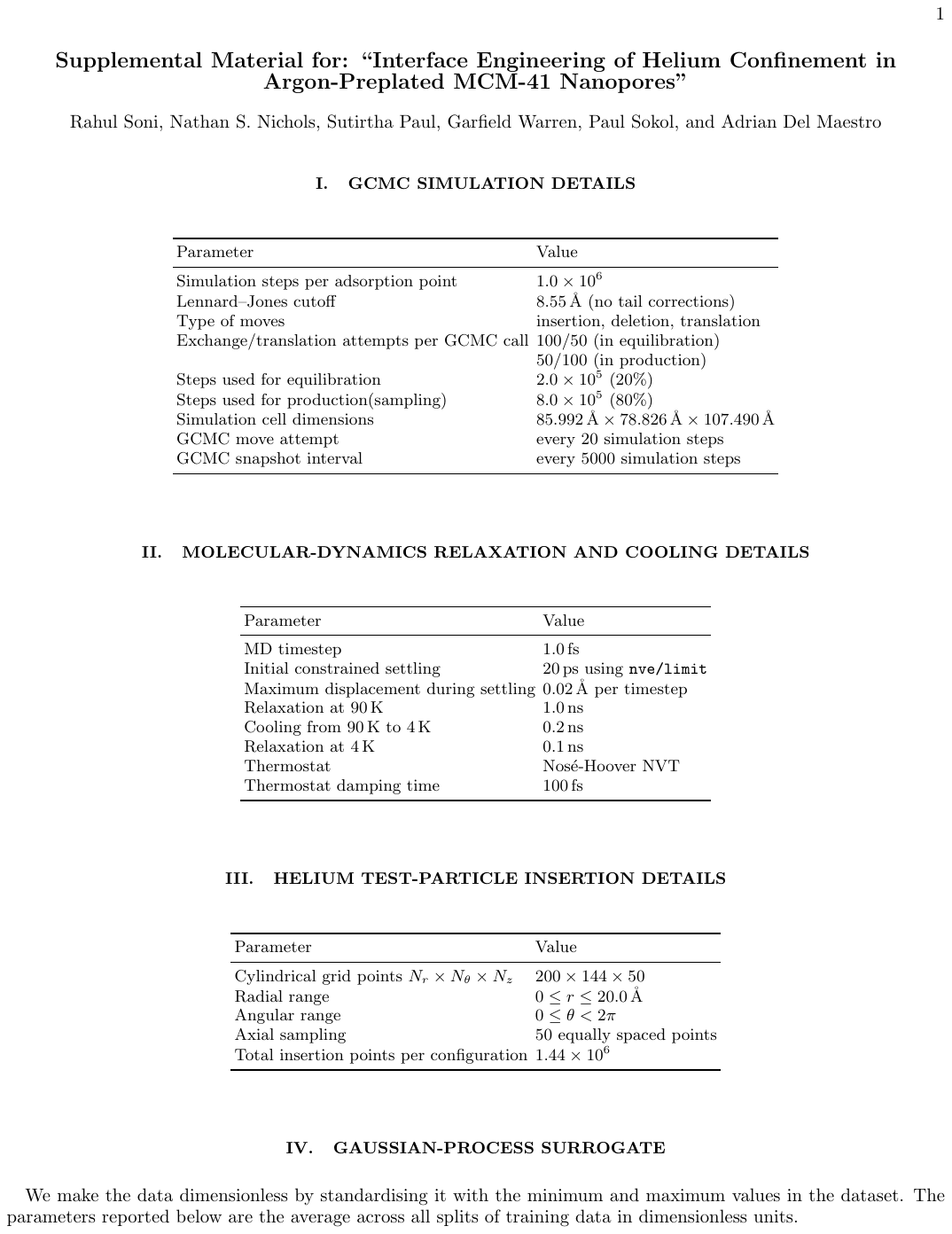}
            }
%\fi

\end{document}